\documentclass[a4paper,12pt]{article}

\usepackage{amsmath,amssymb,amsfonts,epsfig,graphicx,colordvi}

\usepackage{colordvi}

	\newcommand{\state}[2]{|{#1}\rangle_{#2}}
	\newcommand{\nstate}[2]{g^{({#2})}_{|{#1}\rangle}}
	\newcommand{\eq}[1]{$\displaystyle {#1}$} 

\title{Comparison  of the  ferromagnetic Blume-Emery-Griffiths model and 
the AF spin-1 longitudinal Ising  model at low temperature} 

\author{M.T. Thomaz$^{1}\footnote{Corresponding author: mtt@if.uff.br}$ 
and E.V. Corr\^ea Silva$^{2}$
\vspace{0.25cm} \\
\small\it $^{1}$Instituto de F\'{\i}sica, Universidade Federal Fluminense,\\ 
\small\it Av. Gal. Milton Tavares de Souza s/n$^{\textit o}$, 
CEP 24210-346, Niter\'oi-RJ, Brazil.
\vspace{0.25cm}\\ 
\small\it $^{2}$ Departamento de Matem\'atica, 
F\'{\i}sica  e Computa\c{c}\~ao, Faculdade de Tecnologia, \\ 
\small\it Universidade do Estado do Rio de Janeiro. 
Rodovia Presidente Dutra km 298 s/n$^{\textit o}$, \\ 
\small\it P\'olo Industrial, CEP 27537-000,  Resende-RJ, Brazil.
}

\begin{document}

\maketitle

\begin{abstract}
We derive the exact Helmholtz free energy (HFE) of the 
standard and staggered one-dimensional Blume-Emery-Griffiths 
(BEG) model in the presence of an external longitudinal 
magnetic field. We discuss in detail the thermodynamic
behavior of the ferromagnetic version of the model, 
which exhibits magnetic field-dependent plateaux in the 
$z$-component of its magnetization at low temperatures.
We also study the behavior of  its specific heat and 
entropy, both per site, at finite temperature. The degeneracy 
of the ground state, at $T=0$, along the lines that 
separate distinct phases in the phase diagram of the 
ferromagnetic BEG model is calculated, extending the study of
the phase diagram of the spin-1 antiferromagnetic (AF)
Ising model in {\linebreak}[S.M. de Souza and M.T. Thomaz, 
J. of Mag. and Mag. Mat.  {354} (2014) 205]. We explore the 
implications of the equality of phase diagrams, at $T=0$, 
of the ferromagnetic BEG model with $\frac{K}{|J|} = -2$  
and of the spin-1 AF Ising model for 
$\frac{D}{|J|} > \frac{1}{2}$.
\end{abstract}

\vfill


\newpage


\section{Introduction}  \label{sec_1}

For a long time simple 1-D spin models have been used as toy models 
for a better understanding of real systems with coupled spins. 
Experimental verification of the results derived from 
those toy spin models is difficult, given the complexity of 
real spin systems for any spatial dimension. The development 
of optical devices permitted the simulation of a few 1-D spin
models in arrays of cold atoms. In 2011 Simon {\it et al}.\cite{simon} 
simulated the 1-D spin-$1/2$  Ising model in the presence of 
a magnetic field with longitudinal and transverse components 
at low temperature. Such possibility encourage us to explore 
the thermodynamic characteristics of 1-D models.

Recently one of the authors applied the transfer matrix  
method\cite{kramers1,kramers2,baxter} to the calculation of
the exact thermodynamics of the 1-D spin-1 Ising model, with
single-ion anisotropy term, in the presence of an external
longitudinal magnetic field\cite{JMMM2014}.  The present work 
extends that discussion to the classical 1-D Blume-Emery-Griffiths (BEG) 
model\cite{blume} with external longitudinal magnetic field.
This model is classical and its exact thermodynamics can 
also be derived by the transfer matrix method.
The presence of an extra term with respect to the 1-D
Ising model with single-ion anisotropy term can modify 
the behavior of the quantum chain, mainly its phase 
diagram at $T=0$. In the present article we study 
the thermodynamics of the one-dimensional BEG model in 
the presence of an external longitudinal magnetic field.
The phase diagram of the model is discussed in detail 
for the ferromagnetic case, and for two different regions 
of the parameter $\frac{K}{|J|}$, and complemented by the 
discussion on the phase diagram of the spin-1 AF Ising 
model\cite{JMMM2014}.

In section \ref{sec_2} we present the Hamiltonian of the
standard BEG model in the presence of an external 
longitudinal magnetic field. We show the relation between 
the Hamiltonians of the standard and staggered versions 
of this model, to be used in relating their thermodynamics.
In section \ref{sec_3} we discuss the phase diagram, at 
$T=0$, of the ferromagnetic BEG model. Its thermodynamics 
is presented in subsection \ref{sub_3.1} through the 
behavior of three thermodynamic functions per site: the 
$z$-component of the magnetization, the specific heat 
and the entropy. The entropy per site along each line 
that separates distinct phases in the diagram of
the ferromagnetic BEG model is calculated at $T=0$.
In section \ref{sec_4} we compare the three previous 
thermodynamic functions of the ferromagnetic BEG model
with $\frac{K}{|J|}= -2$ and the spin-1 AF Ising model 
at very low temperature. We also extend  the discussion
on the phase diagram of the spin-1 AF  model in 
Ref.\cite{JMMM2014} in order to include the degeneracy of 
the ground state of the model at $T=0$. Our conclusions 
are presented in section \ref{sec_5}. In Appendix \ref{Apend_A}
we present the main steps to calculate the exact Helmholtz 
free energy (HFE) of the standard and staggered one-dimensional 
BEG model for arbitrary values of the parameters.
The states and energies of the dimers present in 
the spin configurations of the chain are shown
in Appendix \ref{Apend_B}. In Appendix \ref{Apend_C}
we have the ground states of the BEG model in the presence 
of an external longitudinal magnetic field and their 
respective energies. Finally in Appendix \ref{Apend_D} we
show  how to calculate the degeneracy of the ground 
state along the lines that separate the different phases 
along the diagrams of the chain models at $T=0$.


\section{The Hamiltonian and HFE of the 1-D Blume-Emery-Griffiths
model with a longitudinal magnetic field}  \label{sec_2}

Eq.(5) of Ref.\cite{JMMM2014},
\begin{eqnarray} \label{2.1}
{\bf H}_I^{S=1} (J, h, D) = \sum_{i=1}^{N} \; \;  \left[ J S_i^z S_{i+1}^z  
- h S_i^z - h S_{i+1}^z + D (S_i^z)^2 + D (S_{i+1}^z)^2 \right],  
\end{eqnarray}
 
\noindent is the Hamiltonian of the one-dimensional classical spin-1 Ising model
with the single-ion anisotropy term with the crystal field $D$, 
the Blume\cite{blume1966} -Capel\cite{capel} model, in the
presence of an external longitudinal ($z$-axis) magnetic field $h$, symmetrized in
the nearest neighbours. Here, $S_i^z$ is the $z$-component  of the
spin-1 operator in the $i$-th site ($|\vec{S}|^2 = 2$), and $J$ is 
the exchange strength. For $J>0$ we have the anti-ferromagnetic (AF) 
version of the model, whereas for $J<0$ ferromagnetic version is obtained.
We assume that $h \ge 0$, that $D$ may have any real value, and that
the chain has $N$ sites and it is periodic, i.e. $S_{N+1}^z = S_1^z$.
In this paper we use natural units $e= m= \hbar =1$. 

Adding the term \eq{-K (S_i^z)^2 (S_{i+1}^z)^2} to Hamiltonian  (\ref{2.1}), 
with \eq{K \in \mathbb{R}}, yields  the Hamiltonian of the Blume-Emery-Griffiths 
(BEG)  model\cite{blume,rys}
\begin{eqnarray} \label{2.2}
{\bf H}_{BEG} (J, h, D, K) = \sum_{i=1}^{N} \; \;   [ J S_i^z S_{i+1}^z  
   &-& h S_i^z - h S_{i+1}^z + D (S_i^z)^2 + D (S_{i+1}^z)^2 
            \nonumber    \\
    &-&  K (S_i^z)^2 (S_{i+1}^z)^2] .        
\end{eqnarray} 

\noindent  This Hamiltonian also satisfies the
periodic condition. (The Hamiltonian (1) with $L=0$
in Ref.\cite{blume} describes the BEG model in a non-symmetrized
form.)

Ref.\cite{JMMM2014} discusses at length which quantum 
state(s) is (are) favored  by each term on the r.h.s. of 
Hamiltonian (\ref{2.1}), regarding the minimization of 
energy; we will not repeat this discussion here. Let $s_i^z$ 
be the eigenvalue of the operator $S_i^z$. In (\ref{2.2}),
the term in $K$ will, for $K>0$, favor the dimer states (i.e., relative
to two neighboring spins) in which $s_i^z = \pm 1$, independently of
their relative orientation (they may be either parallel or
anti-parallel). On the other hand, for $K<0$, the favored dimer states
will be those with at least one null eigenvalue, i.e., $s_i^z = 0$. 
Section \ref{sec_3} will describe how the term in $K$ changes 
the $T=0$ phase diagrams of the classical ferromagnetic spin-1 
Ising models presented in Ref.\cite{JMMM2014}. 

The {\it staggered} BEG model in its symmetrized version reads
\begin{eqnarray} \label{2.3}
{\bf H}_{BEG}^{stag} (J_s, h_s, D_s, K_s)  
&=& \sum_{i=1}^{N} \; \;  \left[ J_s S_i^z S_{i+1}^z  
   - (-1)^i h_s S_i^z - (-1)^{i+1}  h_s S_{i+1}^z
         \right.      \nonumber   \\
    &+&  \left. D_s (S_i^z)^2 + D_s (S_{i+1}^z)^2 
        - K_s (S_i^z)^2 (S_{i+1}^z)^2 \right].  
\end{eqnarray}

\noindent This Hamiltonian will also be subject to the 
spatial periodic condition. We assume that the 
chain has an even number of sites, so $N= 2M$, in 
which $M\in \mathbb{N}$.

The mapping $S_i^z \rightarrow (-1)^i S_i^z$ in 
Hamiltonian (\ref{2.3}) yields the relation between the 
standard and staggered BEG models
\begin{eqnarray}   \label{2.4}
{\bf H}_{BEG}^{stag} (J_s, h_s, D_s, K_s) = {\bf H}_{BEG} ( -J, h, D, K);
\end{eqnarray}

\noindent hence they have the same thermodynamics 
if $J_s = - J$, $h_s = h$, $D_s = D$
and $K_s = K$. The ferromagnetic staggered BEG model ($J_s < 0$) 
has the same thermodynamics as the AF standard
BEG model ($J>0$). The AF staggered BEG model ($J_s > 0$)  
has the same behavior as the ferromagnetic 
standard BEG model ($J<0$) at any finite temperature.

From now on we will restrict our discussion to the
thermodynamics of the standard Hamiltonian (\ref{2.2}) of the BEG 
model. The thermodynamic behavior of the staggered BEG
models at finite temperature can be obtained from the 
corresponding standard models by using (\ref{2.4}).

In appendix \ref{Apend_A} we show the calculation of the 
exact expression of the Helmholtz free energy (HFE) of the 
ferromagnetic and AF BEG models in the presence of 
a longitudinal magnetic field by the transfer matrix
method\cite{kramers1,kramers2,baxter}, valid at any finite 
temperature $T>0$.  In 1975 Krinsky and Furman\cite{krinsky} 
calculated this thermodynamic function for those BEG models. 
Our expression of the HFE for non-null external longitudinal 
magnetic field $h \not= 0$ written differently  from that of 
Ref.\cite{krinsky}; ours has contributions only from real 
functions of the parameters of Hamiltonian (\ref{2.2}) 
and of \eq{\beta = \frac{1}{k T}}, in which $k$ is
Boltzmann's constant and $T$ is the absolute temperature in kelvin. 
Although the results derived in that appendix are 
valid for both the ferromagnetic and the AF BEG 
models, in the following sections of  this paper the 
discussion is restricted to the ferromagnetic case.


\section{The phase diagram of the ferromagnetic  BEG model at $T=0$}  \label{sec_3}

The Hamiltonian (\ref{2.2}) can be written as the sum of 
Hamiltonians of dimers on 
neighboring sites $i$ and $i+1$, $i\in\{ 1, 2, \cdots, N\}$. 
For the dimer composed of the $(i,i+1)$ sites, we have
\begin{eqnarray}   \label{3.1}
{\bf H}_{i, i+1}^{(D)}  (J, h, D, K) &=& J S_i^z S_{i+1}^z  
   - h S_i^z - h S_{i+1}^z + D (S_i^z)^2 + D (S_{i+1}^z)^2 
            \nonumber    \\
    &-&  K (S_i^z)^2 (S_{i+1}^z)^2. 
\end{eqnarray}

\noindent The ferromagnetic case corresponds to $J < 0$.

Let  $|s_i^z\rangle_i$ and $s_i^z$ be the eigenstate and eigenvalue, 
respectively, of the $z$-component  of the spin operator at 
$i$-th site, $S_i^z$, so that $S_i^z |s_i^z\rangle_i = s_i^z |s_i^z\rangle_i$, 
with $s_i^z\in\{-1, 0, 1\}$. The energy $\varepsilon_{i, i+1}$ of 
the dimer $(i, i+1)$, described by the state 
$|D\rangle_{i, i+1} = |s_i^z\rangle_1 \otimes |s_{i+1}^z\rangle_{i+1}$ 
is, in units of $|J|$,
\begin{eqnarray}   \label{3.2}
\frac{\varepsilon_{i, i+1}}{|J|} &=& 
\frac{\phantom{...}_{i, i+1}\langle D| {\bf H}_{i, i+1}^{(D)}|D\rangle_{i, i+1} } {|J|} 
           \nonumber     \\
&=& s_i^z s_{i+1}^z - \frac{h}{|J|} (s_i^z + s_{i+1}^z) 
 + \frac{D}{|J|} \left[\rule{0mm}{5mm} (s_i^z)^2 + (s_{i+1}^z)^2 \right]
   - \frac{K}{|J|} (s_i^z)^2  (s_{i+1}^z)^2,
\end{eqnarray}

\noindent with 
$s_i^z, s_{i+1}^z \in \{ 0, \pm 1\}$, and $i\in \{1, 2, \cdots, N\}$.
All the parameters of the Hamiltonian (\ref{2.2}) are 
scaled in units of $|J|$: \eq{\frac{h}{|J|},\ \frac{D}{|J|}} 
and \eq{\frac{K}{|J|}}; correspondingly, the inverse of the temperature
scales as $|J|\beta$.

In appendix \ref{Apend_B} we present the nine possible 
dimer configurations of neighbouring sites in the chain and 
their respective  energy per unit of $|J|$. 
The ground state of the ferromagnetic BEG model 
is composed of dimer states which minimize the
energy at $T=0$.

The value of the parameter \eq{\frac{K}{|J|}} determines the 
general structure of the $T=0$ phase diagram of the 
ferromagnetic BEG model.

\vspace{0.3cm}

\noindent $i$) The case \eq{\frac{K}{|J|} < -1}.

\vspace{0.15cm}

The $T=0$ phase diagram for this case is represented in Fig.\ref{fig_1}a.
It resembles the phase diagram of the classical 
spin-1 AF Ising model with single ion anisotropy term
and external longitudinal magnetic field,
discussed in Ref.\cite{JMMM2014}.  However, the AF N\'eel 
states $|\Psi_0 \rangle_G$  and $|\Psi_0 \rangle_{G^{\prime}}$
[cf. Eqs. (\ref{C.3a}) and (\ref{C.3b}) of \cite{JMMM2014}]
are naturally absent from Fig.\ref{fig_1}a. The ray 
(half-line) $\frac{h}{|J|} = \frac{D}{|J|}$ extending 
from the origin separates the phases $A$ and $E/E^{\prime}$,
whereas the parallel ray
$\frac{h}{|J|} = \frac{D}{|J|} -1 - \frac{K}{|J|}$, 
extending from the point $\left(\frac{D_1}{|J|}, 0\right)$,
in which $\frac{D_1}{|J|} = 1 + \frac{K}{|J|} < 0$, 
separates the phases $E/E^{\prime}$ and $B$.
Correspondingly, the ray $\frac{h}{|J|} = - \frac{D}{|J|}$ 
from the origin separates the phases $A$ and $F/F^{\prime}$
and the parallel ray 
$\frac{h}{|J|} = -\frac{D}{|J|} +1 + \frac{K}{|J|}$
from  $\left(\frac{D_1}{|J|}, 0\right)$
 separates the phases $F/F^{\prime}$ and $C$.
The two rays from the origin and their parallel rays are 
displaced horizontally by $\frac{|D_1|}{|J|}$.
Such displacement increases as the value of $\frac{K}{|J|}$  
decreases.

The chain states corresponding to each phase are described 
in the Appendix \ref{Apend_C} of this paper.
Those states are represented as \eq{|\Psi_0\rangle_X}, in
which \eq{X \in \{A, B, C, E, E^\prime, F, F^\prime\}.} The phases $\{A,
B, C\}$ correspond to the nondegenerate chain states
\eq{|\Psi_0\rangle_A, |\Psi_0\rangle_B} and \eq{|\Psi_0\rangle_C},
respectively [cf. Eqs. (\ref{C.1a})--(\ref{C.1c})],
whereas the phases \eq{\{E/E^\prime, F/F^\prime\}} are twofold
degenerate, corresponding to the pairs of chain states
\eq{(|\Psi_0\rangle_E, |\Psi_0\rangle_{E^\prime})} and
\eq{(|\Psi_0\rangle_F, |\Psi_0\rangle_{F^\prime})}, respectively [cf.
Eqs.(\ref{C.1d})--(\ref{C.1g})]. 

The number of degenerate ground states corresponding 
to the lines and critical points separating the phases
in Fig \ref{fig_1}a at $T=0$ can also be calculated.
Along one such line, we may determine the possible states of neighboring
sites in the chain, which are those dimer states listed in Appendix
\ref{Apend_B} that minimize the energy along that line. The same
guideline can be applied to the critical points ${\cal R}$ and
${\cal T}$. In appendix \ref{Apend_D} the degeneracy of states on
those lines and points  for $T=0$ is detailed.

Let $\Omega_{U \rightleftharpoons V}(T=0)$
be the number of degenerate ground state vectors
along the line that separates two distinct phases $U$  and 
$V$ of the phase diagram in Fig \ref{fig_1}a, excluding the critical points
${\cal R}$ and ${\cal T}$. These degenerate states do not
necessarily satisfy the periodic spatial boundary 
condition. Our aim is calculating the entropy per site, 
at $T=0$, along the lines that separate the distinct 
phases in the phase diagrams of Figs.\ref{fig_1}a and 
\ref{fig_1}b. In Appendix \ref{Apend_D} we show that the entropy
per site along those lines is the same whether we take into 
account the periodic spatial boundary condition or not.
We obtain
\begin{subequations}
\begin{eqnarray}
\Omega_{B \rightleftharpoons C} (T=0) &=&  2,    \label{3.3a} \\
\Omega_{E/E^{\prime} \rightleftharpoons F/F^{\prime}} (T=0) &=& 
 \frac{3}{4} \; 2^{\frac{N+1}{2}} \left(1 + (-1)^{N+1} \right) 
 + 2^{\frac{N}{2}} \; \left(1 + (-1)^{N}\right),
                        \label{3.3b}   \\
\Omega_{A \rightleftharpoons E/E^{\prime}} (T=0) &=& 
 \Omega_{A \rightleftharpoons F/F^{\prime}} (T=0)
  = \Omega_{B \rightleftharpoons E/E^{\prime}} (T=0)
   = \Omega_{C \rightleftharpoons F/F^{\prime}} (T=0)  \nonumber \\
&=&
 \frac{1}{10} \; \left(\frac{1 + \sqrt{5}}{2}\right)^{N} \, (5 + 3 \sqrt{5})
+ \frac{1}{10} \; \left(\frac{1 - \sqrt{5}}{2}\right)^{N} \, (5 - 3 \sqrt{5}), 
             \nonumber    \\
           \label{3.3c}
\end{eqnarray}
\end{subequations}

\noindent in which $N$ is the total number of sites
in the chain. We are assuming that $N$ is even: 
$N = 2M$, in which $M \in \mathbb{N}$. Moreover, the
number of degenerate ground state vectors at $T=0$,  at the 
critical points ${\cal R}$ and ${\cal T}$ is given by
\begin{subequations}
\begin{eqnarray}
	\Omega_{{\cal R}} 
(T=0) &=& 3 \times 2^{N-1},\ \mbox{and}  \label{3.4a}  \\
	\Omega_{{\cal T}}
 (T=0) &=&  \frac{1}{3}  
	\left(2^{N+2} - (-1)^{N}\right) . 
         \label{3.4b}
\end{eqnarray}
\end{subequations}

\noindent Again $\Omega_{{\cal R}}$ and $\Omega_{{\cal T}}$ 
are the total number of degenerate ground states at these critical 
points, including the states that do not satisfy the periodic
spatial boundary condition.

Except for the line separating the phases $B$ and $C$ in the
phase diagram of Fig.\ref{fig_1}a, all other lines are highly degenerate. 
The results (\ref{3.3a})-(\ref{3.4b}) do not depend on the 
particular value and sign of the  exchange strength $J$; rather, they 
depend only on which dimer configurations yield the minimum energy for 
the parameter scenario along each line. Consequently, a comparison 
of the phase diagram in Fig.\ref{fig_1}a and the phase diagram 
shown in Fig.1b of Ref.\cite{JMMM2014} for the AF spin-1 model 
with single ion-anisotropy and external longitudinal magnetic field
shows that the phases and the lines separating them are the 
same. Moreover, those lines have the same degeneracy in both 
diagrams.

\vspace{0.3cm}

\noindent $ii$) The case \eq{\frac{K}{|J|} \ge -1}.

\vspace{0.15cm}

The $T=0$ phase diagram of the ferromagnetic BEG model in this 
case is shown in Fig.\ref{fig_1}b. The lines
$\frac{h}{|J|} = \mp \frac{1}{2} \pm \frac{D}{|J|} \mp \frac{K}{2|J|}$,
separate the phases $A$ and $B$, and the phases $A$ and 
$C$, respectively. All phases $A$, $B$ and $C$ are nondegenerate, 
and they are described by the chain state vectors
 $|\Psi_0\rangle_A$, $|\Psi_0\rangle_B$ and
$|\Psi_0\rangle_C$, respectively.

Appendix \ref{Apend_B} shows the nine possible dimer states 
and their corresponding energies. Along 
the horizontal line that separates the phases 
$B$ and $C$ in Fig.\ref{fig_1}b to the point $\frac{D_2}{|J|}$, 
the energies of the dimers $|D^{(B)}\rangle_{i,i+1}$ and 
$|D^{(C)}\rangle_{i,i+1}$ [cf. (\ref{B.2}) and (\ref{B.3})] 
are the same and  correspond to the minimum value of energy 
out of the nine possibilities. 
The chain ground states that can be constructed from 
$|D^{(B)}\rangle_{i,i+1}$ and $|D^{(C)}\rangle_{i,i+1}$
along that line are $|\Psi_0\rangle_B$ and 
$|\Psi_0\rangle_C$ [cf. (\ref{C.1b}) and (\ref{C.1c}) 
of Appendix \ref{Apend_C}]. The degree of degeneracy of 
the ground state along this line is equal to $2$.

Likewise, along the line separating the phases $A$ and $B$ in
Fig.\ref{fig_1}b, the dimer states with minimum energy are 
$|D^{(A)}\rangle_{i,i+1}$ and $|D^{(B)}\rangle_{i,i+1}$ 
[cf. (\ref{B.1}) and (\ref{B.2})], and from those the 
chain ground states correspond to $|\Psi_0\rangle_A$ and 
$|\Psi_0\rangle_B$ [cf. (\ref{C.1a}) and (\ref{C.1b}).
Hence the degree of degeneracy of the ground state along 
this line is also equal to 2.

A similar argument holds for the line separating the phases 
$A$ and $C$; its degeneracy is also equal to 2.

\vspace{0.3cm}

The existence or not of an exponentially growing degeneracy 
of the ground states along the separation lines in the phase 
diagrams of Figs.\ref{fig_1}a and \ref{fig_1}b, at $T=0$, determines
the thermodynamic behavior of the ferromagnetic BEG model.


\subsection{Thermodynamic behavior of the ferromagnetic BEG
                model}  \label{sub_3.1}

In this subsection we discuss three thermodynamic
functions {\it per site} of the ferromagnetic ($J<0$) BEG model: the $z$-component of 
the magnetization,
\begin{subequations}
\begin{eqnarray}   \label{3.1.1a}
	{\cal M}_z (J, h, D, K; \beta) = - \frac{1}{2} \frac{\partial {\cal W}}{\partial h},
	\end{eqnarray}
	
\noindent the specific heat

\begin{eqnarray}    \label{3.1.1b}
	{\cal C}(J, h, D, K; \beta) 
	   = -\beta^2 \frac{\partial^2 [\beta {\cal W}]}{\partial \beta^2},
	\end{eqnarray}
	
\noindent and the entropy

\begin{eqnarray}   \label{3.1.1c}
	{\cal S}(J, h, D, K; \beta) =  \beta^2 \frac{\partial {\cal W}}{\partial \beta},
	\end{eqnarray}

\end{subequations}

\noindent in which ${\cal W} (J, h, D, K; \beta)$ is the 
HFE of the model. Its exact expression at finite temperature 
$T$ (in kelvin) can be found in appendix \ref{Apend_A}.

In the following we let $J= -1$; the remaining parameters 
in the Hamiltonian (\ref{2.2}) are in units of $|J|$: 
$\frac{h}{|J|}$, $\frac{D}{|J|}$  and $\frac{K}{|J|}$. The 
inverse of temperature $\beta$ is scaled as $|J| \beta$  
in those functions. Our aim is verifying  the effect of the 
degeneracy along the separating lines in the $T=0$
phase diagrams of Figs.\ref{fig_1}a and \ref{fig_1}b 
on these thermodynamic functions. We will discuss separately 
the regions of the phase diagram in which $\frac{K}{|J|} < -1$  
and $\frac{K}{|J|} \ge -1$.

The case $\frac{K}{|J|} < -1$ is exemplified in Fig.\ref{fig_2}, 
which shows \eq{{\cal M}_z\left(\frac{h}{|J|}\right)} for 
$J = -1$ and $\frac{K}{|J|} = -1.5$.
That corresponds to examining, in the phase diagram of Fig.\ref{fig_1}a, a
vertical straight line in the upper half-plane $\frac{h}{|J|} \ge 0$. Two
scenarios can be qualitatively distinguished.  The first 
scenario is exemplified by Fig.\ref{fig_2}a, which shows
 ${\cal M}_z\left(\frac{h}{|J|}\right)$ 
for $\frac{D}{|J|} = -2$ and $|J| \beta \in \{1.5, 5, 100\}$. 
This figure corresponds to following upwards the vertical line
\eq{L_1} in the phase diagram in Fig.\ref{fig_1}a,  which 
begins {{at the point corresponding to $\frac{D}{|J|} =-2 $ 
and $ h=0$, on the horizontal axis}}.  Along the 
vertical line \eq{L_1}, the degree of degeneracy of the
ground state vector is 2. Fig.\ref{fig_2}a shows that for 
$(J= -1, \frac{K}{|J|} = -1.5$ and
$\frac{D}{|J|} = -2)$, the step-function shape of ${\cal M}_z
\left(\frac{h}{|J|}\right)$ is lost about $|J| \beta = 1.5$.
The second scenario is exemplified by Fig.\ref{fig_2}b, 
which shows  ${\cal M}_z\left(\frac{h}{|J|}\right)$
for $\frac{D}{|J|} = 0.25$ and $|J| \beta \in \{10, 100, 1000\}$.
It corresponds to following upwards the vertical line 
\eq{L_2} in the phase diagram in Fig.\ref{fig_1}a,  
starting from the point corresponding to 
$\frac{D}{|J|} = 0.25$ and $h=0$, on the horizontal axis.
$L_2$ intersects two phase transition lines: from $A$ to $E/E^\prime$ at
$\frac{h}{|J|}= 0.25$ and from $E/E^\prime$ to $B$ at $\frac{h}{|J|} = 0.75$. 
The degeneracies of the ground state vectors along these two transition lines 
are exponential [cf. Eq.(\ref{3.3c})]. Consequently, the shape 
of the curve ${\cal M}_z (\frac{h}{|J|})$ at $|J| \beta =10$ 
is quite distinct from that of the corresponding curve at $|J| \beta = 1000$.
For the latter, the $z$-component of the magnetization
has two plateaux that resemble quite closely a sequence of step-functions.
At $|J| \beta = 10$, however, the plateaux are no longer present. 
That same behavior of ${\cal M}_z\left(\frac{h}{|J|}\right)$ 
for different temperatures is seen in any curve  for 
$\frac{D}{|J|} > \frac{D_1}{|J|}$, in which 
$\frac{D_1}{|J|} = 1 + \frac{K}{|J|}$, and $\frac{K}{|J|} < -1$.

Figs.\ref{fig_3} show  the function 
${\cal M}_z (\frac{h}{|J|})$ for $\frac{K}{|J|}  \ge -1$.
In Figs.\ref{fig_3}a and \ref{fig_3}b we have $J= -1$ (ferro)
and $\frac{K}{|J|}  = 1$. In Fig.\ref{fig_3}a we choose
$\frac{D}{|J|} = -0.5$. The vertical line in the phase
diagram of Fig.\ref{fig_1}b  followed by the argument  
$\frac{h}{|J|}$ of the function ${\cal M}_z$, begins 
at $\frac{D}{|J|} = - 0.5$ with $h=0$ and crosses the 
phase $B$ of this diagram  at $T=0$. 
From Fig.\ref{fig_3}a we see that the function 
${\cal M}_z (\frac{h}{|J|})$ has the  plateau ${\cal M}_z =1$ 
up to $|J| \beta \sim 5$. For $|J| \beta = 1.5$ the curve 
${\cal M}_z (\frac{h}{|J|})$ in Fig.\ref{fig_3}a differs  from
the step-function around $\frac{h}{|J|} \sim 0$. 
One is reminded that the degeneracy  of the ground 
state vectors along the line that separates the phases 
$B$ and $C$ in Fig.\ref{fig_1}b is equal to 2.

In Fig.\ref{fig_3}b we have $\frac{D}{|J|} = 1.8$. The function 
${\cal M}_z (\frac{h}{|J|})$ is plotted for the variable $\frac{h}{|J|}$
varying along the vertical line in the phase diagram \ref{fig_1}b 
with $\frac{D}{|J|} = 1.8$. This vertical line crosses 
the phase $A$ and cut the line that separates the phases 
$A$ and $B$ in the phase diagram \ref{fig_1}b at $\frac{h}{|J|} = 0.8$,
going along the phase $B$. The number of degenerate ground
state vectors along the line separating the phases $A$ and $B$ in the 
phase diagram is equal to 2. In Fig.\ref{fig_3}b the curves 
of the function ${\cal M}_z (\frac{h}{|J|})$ has two plateaux,
${\cal M}_z = 0$ and ${\cal M}_z =1$, for $|J| \beta$ up to 5. It 
looses the step-function form for $|J| \beta \sim 1.5$ (a 
high temperature $T \sim \frac{2 |J|}{3}$). Comparing the 
curves of the $z$-component of the magnetization as a function
of the $\frac{h}{|J|}$  in Figs.\ref{fig_2}a, \ref{fig_3}a 
and \ref{fig_3}b, we verify that the plateaux in those curves 
are present up to a high enough temperature. The common fact 
about these three curves is that the vertical line in the 
phase diagrams of Fig. \ref{fig_1} (for an increasing value 
of $\frac{h}{|J|}$ and a steady value of $\frac{D}{|J|}$)
cut phase transition lines for which the degeneracy of the 
ground state vectors are not exponential.

\vspace{0.3cm}

The entropy per site ${\cal S} (J, h, D, K; \beta)$  can be 
derived from the HFE ${\cal W} (J, h, D, K; \beta)$ of the model
through the relation between these two functions presented
previously, see eq.(\ref{3.1.1c}), or through the 
number of states with energy between $ \bar{E}$ and 
$\bar{E} + \overline{\delta E}$, with 
$\overline{\delta E} \ll \bar{E}$\cite{reif}, 
$\Omega(\bar{E})$, in the thermodynamic limit, 
\begin{eqnarray}   \label{3.1.1}
{\cal S} (J, h, D, K; \beta) = \lim \limits_{N \to \infty}  
       \frac{k}{N} \; 
	      ln(\Omega (\bar{E})).
\end{eqnarray}

\noindent From this point on, we will use a system of units 
in which $k=1$.

Using an algebraic manipulation program we derive from 
the results of appendix \ref{Apend_A}  the temperature
dependence of the entropy per site  of the BEG model 
(ferromagnetic and AF models), valid for $T>0$.  In 
eqs.(\ref{3.3a}) - (\ref{3.4b}) we present the number 
of degenerate ground state vectors along the line that separate 
the different phases in the diagrams \ref{fig_1}a and \ref{fig_1}b, 
at $T=0$, and the critical points ${\cal R}$ and ${\cal T}$.

By varying the inverse of temperature, $|J| \beta$, 
up to $10^7$, we verify numerically that the entropy per site, 
${\cal S} (-1, \frac{h}{|J|}, \frac{D}{|J|}, \frac{K}{|J|}; |J| \beta)$,
has a strong indication  that the limit of this thermodynamic 
function as $T \rightarrow 0$ ($|J| \beta \rightarrow  \infty$) along
the lines that separate the phases in Figs.\ref{fig_1}a 
and \ref{fig_1}b are as follows.

\noindent 1) For  $\frac{K}{|J|}  < -1$  (phase diagram 
\ref{fig_1}a) we have, for the \eq{{B \rightleftharpoons C}}
 transition line,

\begin{subequations}
\begin{eqnarray}   \label{3.1.2a}
\lim \limits_{|J| \beta \to \infty}  
   {\cal S}_{B \rightleftharpoons C} (-1, 0, \frac{D}{|J|}, \frac{K}{|J|}; |J| \beta)
= \lim \limits_{N \to \infty} \; \frac{1}{N}  \; 
            ln(\Omega_{B \rightleftharpoons C} (T=0)) = 0,
\end{eqnarray}

\noindent with $\frac{D}{|J|} < \frac{D_1}{|J|}$ and 
$\frac{D_1}{|J|} = 1 + \frac{K}{|J|}$. For the 
\eq{{{E/E^{\prime} \rightleftharpoons F/F^{\prime}}}} transition line,
\begin{eqnarray}   \label{3.1.2b}
\lim \limits_{|J| \beta \to \infty}  
   {\cal S}_{E/E^{\prime} \rightleftharpoons F/F^{\prime}}
         (-1, 0, \frac{D}{|J|}, \frac{K}{|J|}; |J| \beta)
&=& \lim \limits_{N \to \infty}  \;  \frac{1}{N}  \; 
  ln(\Omega_{E/E^{\prime} \rightleftharpoons F/F^{\prime}} (T=0)) \nonumber  \\
&=& \frac{1}{2} \, ln (2) \approx 0.3466, 
\end{eqnarray}

\noindent with $ \frac{D_1}{|J|} < \frac{D}{|J|} < 0$. For the 
\eq{{B \rightleftharpoons E/E^{\prime}}} transition line,
\begin{eqnarray}   \label{3.1.2c}
\lim \limits_{|J| \beta \to \infty}  
   {\cal S}_{B \rightleftharpoons E/E^{\prime}}
        (-1, \frac{h}{|J|}, \frac{D}{|J|}, \frac{K}{|J|}; |J| \beta)
          &=& \lim \limits_{|J| \beta \to \infty}  
   {\cal S}_{C \rightleftharpoons F/F^{\prime}}
        (-1, \frac{h}{|J|}, \frac{D}{|J|}, \frac{K}{|J|}; |J| \beta)
             \nonumber    \\
&=& \lim \limits_{N \to \infty}  \; \frac{1}{N}  \; 
  ln(\Omega_{B \rightleftharpoons E/E^{\prime}} (T=0)) \nonumber  \\
&=& \lim \limits_{N \to \infty}  \; \frac{1}{N}  \; 
  ln(\Omega_{C \rightleftharpoons F/F^{\prime}} (T=0)) \nonumber  \\
&=& ln\left(\frac{1 + \sqrt{5}}{2}\right) \approx 0.4812, 
\end{eqnarray}

\noindent with $\frac{h}{|J|} = \pm \frac{D}{|J|}  \mp 1 \mp \frac{K}{|J|}$
and $\frac{D}{|J|} > \frac{D_1}{|J|}$. For the 
\eq{{A \rightleftharpoons E/E^{\prime}}} transition line,
\begin{eqnarray}   \label{3.1.2d}
\lim \limits_{|J| \beta \to \infty}  
   {\cal S}_{A \rightleftharpoons E/E^{\prime}} 
      (-1, \frac{h}{|J|}, \frac{D}{|J|}, \frac{K}{|J|}; |J| \beta)
          &=&  \lim \limits_{|J| \beta \to \infty}  
   {\cal S}_{A \rightleftharpoons F/F^{\prime}} 
      (-1, \frac{h}{|J|}, \frac{D}{|J|}, \frac{K}{|J|}; |J| \beta)
          \nonumber   \\
&=& \lim \limits_{N \to \infty}  \; \frac{1}{N}  \; 
  ln(\Omega_{A \rightleftharpoons E/E^{\prime}} (T=0)) \nonumber  \\
&=& \lim \limits_{N \to \infty}  \; \frac{1}{N}  \; 
  ln(\Omega_{A \rightleftharpoons F/F^{\prime}} (T=0)) \nonumber  \\
&=& ln\left(\frac{1 + \sqrt{5}}{2}\right) \approx 0.4812, 
\end{eqnarray}

\noindent with $\frac{h}{|J|} = \pm \frac{D}{|J|}$ and $\frac{D}{|J|} > 0$.

In the multicritical points ${\cal R}$ and ${\cal T}$, at $T=0$, we 
have the limits:
\begin{eqnarray}  \label{3.1.2e}
\lim \limits_{|J| \beta \to \infty}  
   {\cal S}_{{\cal R}} (-1, 0, \frac{D_1}{|J|}, \frac{K}{|J|}; |J| \beta)
 &=& \lim \limits_{|J| \beta \to \infty} 
    {\cal S}_{{\cal T}} (-1, 0, 0, \frac{K}{|J|}; |J| \beta)
             \nonumber    \\
&=& \lim \limits_{N \to \infty}  \; \frac{1}{N}  \; 
  ln(\Omega_{{\cal R}} (T=0)) \nonumber  \\
&=& \lim \limits_{N \to \infty}  \; \frac{1}{N}  \; 
  ln(\Omega_{{\cal T}} (T=0)) \nonumber  \\
&=& ln (2) \approx 0.6931,
\end{eqnarray}

\noindent where $\frac{D}{|J|} =  \frac{D_1}{|J|} = 1 + \frac{K}{|J|}$ 
(point ${\cal R}$) or $\frac{D}{|J|} = 0$ (point ${\cal T}$). 

\end{subequations}

\vspace{0.3cm}

\noindent 2) For  $\frac{K}{|J|} \ge 1$ (phase diagram \ref{fig_1}b)
we also vary the inverse of the temperature, $|J| \beta$, 
in the interval $[0, 10^8]$, and the results strongly 
indicate that
\begin{eqnarray}    \label{3.1.3}
\lim \limits_{|J| \beta \to \infty}  
   {\cal S} (-1, \frac{h}{|J|}, \frac{D}{|J|}, \frac{K}{|J|}; |J| \beta) = 0,
\end{eqnarray}

\noindent for any value for the parameters $\frac{h}{|J|}$, $\frac{D}{|J|}$,
when $\frac{K}{|J|} \ge -1$.

\vspace{0.3cm}

Fig.\ref{fig_4} represents the entropy per site as a function of 
$\frac{D}{|J|}$  for $\frac{K}{|J|} = -1.5$ at $|J| \beta = 10^3$. The 
twin peaks in the curve correspond to the points ${\cal R}$
($\frac{D}{|J|} = -0.5$) and ${\cal T}$ ($\frac{D}{|J|} = 0$)
in the phase diagram \ref{fig_1}b. Since $T>0$ for this picture, 
the curve is continuous everywhere. We verify that at low 
temperature the points ${\cal R}$ and ${\cal T}$ are well 
featured in the curve ${\cal S} \times \frac{D}{|J|}$.

\vspace{0.5cm}

The relation 
\eq{\frac{{\cal C} (T)}{T} = \frac{\partial {\cal S} (T)}{\partial T}}
connects  the specific heat per site, in units of the 
temperature, and the variation of the entropy per site
with the temperature. In what follows we will
 discuss the behavior of the specific heat per site 
in the two regions of the parameter $\frac{K}{|J|}$:
$\frac{K}{|J|} < -1$ and $\frac{K}{|J|} \ge -1$.

For $\frac{K}{|J|} < -1$ and $\frac{D}{|J|} < 1 + \frac{K}{|J|}$, 
the maximum value of the specific heat per site, as function 
of $\frac{h}{|J|}$ at $|J| \beta = 10 ^2$, is of order 
$10^{-85}$. For this set of values of the parameters  
$\frac{D}{|J|}$ and $\frac{K}{|J|}$, increasing the variable
 $\frac{h}{|J|}$ corresponds to following
 a vertical line in the phase diagram \ref{fig_1}a that 
does not cut  any line in the diagram that has an exponential 
degeneracy in the ground state at $T=0$.  The maximum value
of ${\cal C}(\frac{h}{|J|})$ at $|J| \beta = 10$ is of order
$10^{-7}$ and at $|J| \beta =2$ the highest value of this 
thermodynamic function is of order $10^{-2}$. For the 
temperature varying two orders of magnitude  the specific 
heat per site varies $10^{83}$ orders of magnitude.

The specific heat per site  with $\frac{K}{|J|} \ge -1$
and $\frac{D}{|J|} \in \mathbb{R}$
has the same behavior as described in the previous paragraph. 
The phase diagram of the chain with $\frac{K}{|J|} \ge -1$
has no line in which the ground state is exponentially 
degenerate at $T=0$.

In Fig.\ref{fig_5} the curve ${\cal C} \times  \frac{h}{|J|}$ 
is drawn with $\frac{K}{|J|} = -1.5$ and $\frac{D}{|J|} = 0.25$.
The variable $\frac{h}{|J|} \ge 0$ follows a vertical line
in the phase diagram \ref{fig_1}a, at $T=0$, that crosses
two lines  that separate the phases $A$ and $E/E^{\prime}$
at $\frac{h}{|J|} = 0.25$, and $E/E^{\prime}$ and $B$ at
$\frac{h}{|J|} = 0.75$. Along  these two lines that separate 
theses phases in diagram \ref{fig_1}a, the ground state vectors
are exponentially degenerate at $T=0$. In Figs.\ref{fig_5}a
and \ref{fig_5}b the curve ${\cal C} (\frac{h}{|J|})$ is plotted 
at $|J| \beta = 10^3$, but in two intervals of the variable
$\frac{h}{|J|}$ where this thermodynamic function is non null 
at this temperature. In Fig.\ref{fig_5}c the specific heat per site 
is drawn at $|J| \beta = 10$.

We verify from the Figs.\ref{fig_5} that the maximum value
of the function ${\cal C} (\frac{h}{|J|})$ at $|J| \beta = 10^3$
is of the same order of magnitude as at $|J| \beta =10$. The curve
of the specific heat per site as a function of $\frac{h}{|J|}$ at
different temperatures, has the same behavior as described in 
Figs.\ref{fig_5} when $\frac{K}{|J|} < -1$ and 
$\frac{D}{|J|} > 1 + \frac{K}{|J|}$.


\section{Ferromagnetic BEG model at $\frac{K}{|J|} = -1  
\times$ AF Ising model at low temperature}     \label{sec_4}

Ref.\cite{JMMM2014}  presents the phase diagram of the 
spin-1 AF Ising model, with single-ion anisotropy term, in the 
presence of a longitudinal external magnetic field, at $T=0$
(see Fig.\ref{fig_6}a). In Fig.\ref{fig_6}b we have the 
phase diagram, also at $T=0$, of the ferromagnetic BEG
model with $\frac{K}{|J|} = -2$, in the presence of an 
external longitudinal magnetic field. Comparing the phase 
diagrams of these two models, we verify that for 
$\frac{D}{|J|} > \frac{1}{2}$  the two models have the same 
phases at $T=0$. The line $l_1$ in the diagram \ref{fig_6}b 
corresponds to the vertical line $\frac{D}{|J|} = \frac{1}{2}$.
(Having the same phases for $\frac{D}{|J|} > \frac{1}{2}$, 
however, is not a sufficient condition for the existence of 
a transformation that maps one model onto the other.)

In subsection \ref{sub_3.1} we discussed the degeneracy
of the ground state along the phase separating lines 
in the diagram of the ferromagnetic BEG model, at $T=0$, 
concluding that the results (\ref{3.3a})-(\ref{3.3c}) are 
independent of the sign of $J$. A analogous discussion for 
the spin-1 AF Ising model (see Fig.\ref{fig_6}a) has 
{\it not} been done in Ref.\cite{JMMM2014}, though.

The independence of the number of degenerate ground states 
with respect to the sign of the exchange strength $J$, 
calculated in subsection \ref{sub_3.1}, permits us to 
affirm that the degeneracy of the ground state of the spin-1
AF Ising model along the lines $\frac{h}{|J|} = \pm  \frac{D}{|J|}$ 
with $\frac{D}{|J|} > \frac{1}{2}$ in diagram \ref{fig_6}a, at $T=0$,
is equal to the result (\ref{3.3c}). Again we are also including the 
states that do not satisfy the periodic spatial boundary condition.
The degeneracy of the ground state of the spin-1 AF Ising model,
at $T=0$, along the line $\frac{h}{|J|} = \pm 1 \pm \frac{D}{|J|}$,
with $\frac{D}{|J|} >0$, in the phase diagram \ref{fig_6}a
is equal to result (\ref{3.3b}), under the same situation 
on the spatial boundary condition.

There are phase transitions at $T=0$ in the phase diagram 
\ref{fig_6}a of the spin-1 AF Ising model that are absent in
the ferromagnetic BEG model with $\frac{K}{|J|} = -2$. In the 
following we present the total number of ground states along
the lines that separate the phases in the spin-1 AF Ising model.
These degenerate ground states do not necessarily satisfy 
periodic spatial boundary conditions:
\begin{subequations}

\begin{eqnarray}
\Omega_{A \rightleftharpoons G/G^{\prime}}^{AF} (T=0) &=&  3, \label{4.1a} \\
\Omega_{G/G^{\prime} \rightleftharpoons E/E^{\prime}}^{AF} (T=0) &=& 
\Omega_{G/G^{\prime} \rightleftharpoons F/F^{\prime}}^{AF} (T=0)  \nonumber \\
  &=& 2^{\frac{N+2}{2}} \, \left( \frac{1+ (-1)^N}{2} \right)
  + (2^{\frac{N+1}{2}} + 2^{\frac{N-1}{2}}) \, \left( \frac{1+ (-1)^{N+1}}{2} \right),
          \label{4.1b}    \\
\Omega_{B \rightleftharpoons G/G^{\prime}}^{AF} (T=0) &=& 
  \Omega_{C \rightleftharpoons G/G^{\prime}}^{AF} (T=0)    \nonumber   \\
  &=& \left( 1+ \frac{2 \sqrt{5}}{5} \right) \, \left( 1 + \frac{\sqrt{5}}{2} \right)^{N-1}
 + \left( 1- \frac{2 \sqrt{5}}{5} \right) \, \left( 1 - \frac{\sqrt{5}}{2} \right)^{N-1}.
                 \label{4.1c}
\end{eqnarray}
\end{subequations}

The number of degenerate ground states at the critical points
$({\cal P}, {\cal P}^{\prime})$ in the phase diagram \ref{fig_6}a is
\begin{eqnarray}    \label{4.2}
\Omega_{{\cal P}, {\cal P}^{\prime}}^{AF} &=& \Omega_{{\cal P}^{\prime}}^{AF}
        \nonumber    \\
&=& \frac{2^{N+2}}{3} + \frac{(-1)^{N+1}}{3}.
\end{eqnarray}

We do not show here (the lengthy expression of) the number 
of degenerate ground states  at the critical points 
$({\cal Q}, {\cal Q}^{\prime})$ in the phase diagram \ref{fig_6}a 
of the spin-1 AF Ising model, at $T=0$. Its calculation 
has been done with the help of a computer algebra system; 
the interested reader may contact the authors for an ASCII 
file with that expression.

\vspace{0.3cm}

The entropy per site of these lines in the phase diagram of the 
spin-1 AF Ising model in Fig.\ref{fig_6}a, in the thermodynamic limit
($N \rightarrow \infty$), is
\begin{subequations}

\begin{eqnarray}
&\lim \limits_{|J| \beta \to \infty}& \!  
   {\cal S}_{B \rightleftharpoons G/G^{\prime}}^{AF} 
   (1, \frac{h}{|J|}, \frac{D}{|J|} = \frac{1}{2}, \frac{K}{|J|} = 0; |J| \beta)  =
                  \nonumber   \\
&=& \lim \limits_{|J| \beta \to \infty}  
   {\cal S}_{C \rightleftharpoons G/G^{\prime}}^{AF} 
   (1, \frac{h}{|J|}, \frac{D}{|J|} = \frac{1}{2}, \frac{K}{|J|} = 0; |J| \beta)
                    \nonumber     \\
  &=& ln \left(\frac{1 + \sqrt{5}}{2} \right)  \approx 0.4812  \label{4.3a} \\
&=& \lim \limits_{|J| \beta \to \infty}  
   {\cal S}_{A \rightleftharpoons E/E^{\prime}} 
   ( -1, \frac{h}{|J|}, \frac{D}{|J|}, \frac{K}{|J|}; |J| \beta),     \label{4.3b}
\end{eqnarray}

\end{subequations}

\noindent with $\frac{K}{|J|} < -1$ and $\frac{D}{|J|}  > 0$. 
We also have
\begin{subequations}

\begin{eqnarray}
&\lim \limits_{|J| \beta \to \infty}& \!  
   {\cal S}_{E/E^{\prime} \rightleftharpoons G/G^{\prime}}^{AF} 
   (1, \frac{h}{|J|}, \frac{D}{|J|}, \frac{K}{|J|} = 0; |J| \beta)  =
           \nonumber    \\
&=& \lim \limits_{|J| \beta \to \infty}  
   {\cal S}_{F/F^{\prime} \rightleftharpoons G/G^{\prime}}^{AF} 
   (1, \frac{h}{|J|}, \frac{D}{|J|}, \frac{K}{|J|} = 0; |J| \beta)  
     = \frac{1}{2} \, ln(2) \approx 0.3466     \label{4.4a} \\
&=& \lim \limits_{|J| \beta \to \infty}   
   {\cal S}_{E/E^{\prime} \rightleftharpoons F/F^{\prime}} 
   (-1, \frac{h}{|J|}, \frac{D}{|J|}, \frac{K}{|J|}; |J| \beta).
   \end{eqnarray}

\end{subequations}

\noindent In the phase diagram of the spin-1 AF Ising model we
have $ \frac{1}{2} < \frac{|h|}{|J|} < 1$ and 
$0 < \frac{D}{|J|} < \frac{1}{2}$. In the phase diagram of the 
ferromagnetic BEG model, at $T=0$, we have $\frac{K}{|J|} < -1$.

The entropy per site of the critical points $( {\cal P}, {\cal P}^{\prime})$, 
in the thermodynamic limit ($N \rightarrow \infty$) at $T=0$, is
\begin{subequations}

\begin{eqnarray}
&\lim \limits_{|J| \beta \to \infty} & \!  {\cal S}_{{\cal P}^{AF}, {\cal P}^{\prime}} 
 (1, \frac{h}{|J|} = \pm 1, \frac{D}{|J|} = 0, \frac{K}{|J|} = 0; |J| \beta) 
      = ln (2) \approx 0.6931   \label{4.5a} \\
&=&  \lim \limits_{|J| \beta \to \infty}   
   {\cal S}_{E/E^{\prime} \rightleftharpoons F/F^{\prime}} 
   (-1, \frac{h}{|J|}, \frac{D}{|J|}, \frac{K}{|J|}; |J| \beta)  \nonumber \\
&=& \lim \limits_{|J| \beta \to \infty} 
    {\cal S}_{\cal T}  (-1, \frac{h}{|J|} = 0, \frac{D}{|J|} = 0, \frac{K}{|J|}; |J| \beta)
 = \lim \limits_{|J| \beta \to \infty} 
    {\cal S}_{\cal R}  (-1, \frac{h}{|J|} = 0, \frac{D_1}{|J|}, \frac{K}{|J|}; |J| \beta),
        \nonumber   \\
&&            \label{4.5b}
\end{eqnarray}

\end{subequations}

\noindent where $\frac{D_1}{|J|} = 1 + \frac{K}{|J|}$, with $\frac{K}{|J|} < -1$.
We also have at the critical points $( {\cal Q}, {\cal Q}^{\prime})$
\begin{eqnarray}   \label{4.6}
\lim \limits_{|J| \beta \to \infty} & \!  {\cal S}_{{\cal Q}, {\cal Q}^{\prime}}^{AF} 
 (1, \frac{h}{|J|} = \pm \frac{1}{2}, \frac{D}{|J|} = \frac{1}{2}, \frac{K}{|J|} = 0; |J| \beta)
     \approx 0.5886. 
\end{eqnarray}

\noindent The quantity
$\lim \limits_{|J| \beta \to \infty} {\cal S}_{{\cal Q}, {\cal Q}^{\prime}}^{AF}$
can be obtained with arbitrary precision.

The previous information about the degeneracy of the 
ground state of the spin-1 Ising model at $T=0$ and along 
the lines that separate the phases in the diagram \ref{fig_6}a, 
complements the information of Ref.\cite{JMMM2014} 
about the phase diagram of the spin-1 AF Ising model, with 
single-ion anisotropy term and in the presence of an 
external longitudinal magnetic field, at $T=0$.

Let us examine in what follows the consequences of 
having the same phase diagram at $T=0$ for 
$\frac{D}{|J|} > \frac{1}{2}$ (see the diagrams of 
Fig.\ref{fig_6}), in the presence of an external 
longitudinal magnetic field, for both the ferromagnetic 
BEG model (with $\frac{K}{|J|} = -2$) and the spin-1 AF
Ising model with single-ion anisotropy term 
(with $\frac{K}{|J|} =0$), 

At very low temperature, the contribution to the thermodynamic 
functions of a model comes mainly from its ground 
state, degenerate or not. In the following we compare 
the $z$-component of the 
magnetization, specific heat and entropy {\it per site}
[in eqs.(\ref{3.1.1a}),  (\ref{3.1.1b}) and (\ref{3.1.1c}), 
respectively] of both models at $|J| \beta = 500$ and 
$|J| \beta = 100$. At the latter temperature, we expect 
that the excited states of the models
give contribution to the thermodynamic functions due to the
exponential degeneracy of these states in each model.

The big difference between the ferromagnetic BEG model, under
consideration in this section, and the spin-1 AF Ising model
is the sign of the exchange strength $J$. In the 
former (latter) model we have $J= -1$ ($J= 1$). We 
notice from the Hamiltonians  (\ref{2.1}) and (\ref{2.2})  
that in order to compare the contribution
of the $J$-term to the partition function of the model and the
contributions from the external longitudinal magnetic field  
and the single-ion anisotropy term, we have to compare the 
values of the parameters: $J$, $2h$ and $2D$.

For a fixed value of $\frac{D}{|J|} > \frac{1}{2}$, 
we have varied the external longitudinal magnetic 
field in whole interval, that is, 
$ \frac{h}{|J|} \in [0, \infty)$ and have compared the 
three thermodynamic functions mentioned previously.

Along the vertical line in the phase diagrams of 
Fig.\ref{fig_6}, with fixed value of $\frac{D}{|J|}$, 
we have an interval of $\frac{h}{|J|} \in [0, \frac{D}{|J|})$ 
for which ${\cal M}_z$ vanishes  at $|J| \beta =500$ 
and $|J|  \beta = 100$. In order to avoid any singular 
point in this comparison we define the difference of 
these functions as
\begin{eqnarray}   \label{4.7}
Diff (M_z)_{Fe}^{AF} \left(\frac{h}{|J|}, \frac{D}{|J|}; 
     |J| \beta\right) &\equiv&
{\cal M}_z^{Fe} \left(-1, \frac{h}{|J|}, \frac{D}{|J|}, \frac{K}{|J|}=-2; 
        |J| \beta\right)  \nonumber   \\
&-& 
{\cal M}_z^{AF} \left(1, \frac{h}{|J|}, \frac{D}{|J|}, \frac{K}{|J|}=0; 
            |J| \beta\right).
\end{eqnarray}

Fig.\ref{fig_7}a shows the magnetization difference (\ref{4.7}) as a
function of $\frac{h}{|J|}$ for $\frac{D}{|J|} = 0.51$ and $|J| \beta =
500$, in an interval about $\frac{h}{|J|} \sim 0.51$ with amplitude of
order $10^{-10}$, in which the transition $A \rightleftharpoons
E/E^{\prime}$ occurs. For $\frac{h}{|J|} \gtrsim 0.511$, 
the difference (\ref{4.7}) of the functions ${\cal M}_z$ is a 
monotonically decreasing function, where we have:
\begin{subequations}

\begin{eqnarray}
Diff (M_z)_{Fe}^{AF} \left(\frac{h}{|J|} = 0.52, \frac{D}{|J|} = 0.51;  
    |J| \beta = 500\right) &\approx & 4.6 \times 10^{-14},   
             \label{4.8a}   \\
Diff (M_z)_{Fe}^{AF} \left(\frac{h}{|J|} = 1, \frac{D}{|J|} = 0.51;  
    |J| \beta = 500\right) &\approx & 1.6 \times 10^{-222},  
             \label{4.8b}   \\
Diff (M_z)_{Fe}^{AF} \left(\frac{h}{|J|} = 1.51, \frac{D}{|J|} = 0.51;  
    |J| \beta = 500\right) &\approx & 3.8 \times 10^{-444}. 
             \label{4.8c} 
\end{eqnarray}

\end{subequations}

One is reminded that at $\frac{h}{|J|} = 1.51$ we have the phase 
transition $B \rightleftharpoons E/E^{\prime}$ in the
phase diagrams of Fig.\ref{fig_6}, at $T=0$.

Keeping $\frac{D}{|J|} = 0.51$ and increasing the
temperature to $|J| \beta =100$, the difference of the
$z$-components of the magnetization of the two models
decreases; it is shown in Fig.\ref{fig_7}b. 
For $\frac{h}{|J|} \sim 0.51$, the value of the parameter
$\frac{2h}{|J|}$ is close to 1 and the ferro- or 
antiferromagnetic nature of the models begins to appear. 
For $\frac{h}{|J|} \gtrsim 0.516$, the  function 
$Diff (M_z)_{Fe}^{AF}$  is a monotonically decreasing 
function of the external magnetic field $\frac{h}{|J|}$ 
and
\begin{subequations}

\begin{eqnarray}
Diff (M_z)_{Fe}^{AF} \left(\frac{h}{|J|} = 1, \frac{D}{|J|} = 0.51;  
    |J| \beta = 100\right) &\approx & 2.5 \times 10^{-45}, 
              \label{4.9a}    \\
Diff (M_z)_{Fe}^{AF} \left(\frac{h}{|J|} = 1.51, \frac{D}{|J|} = 0.51;  
    |J| \beta = 100\right) &\approx & 9.2 \times 10^{-90}. 
              \label{4.9b}    
\end{eqnarray}

\end{subequations}

For $\frac{D}{|J|}=1$ we obtain that the difference (\ref{4.7})
of the $z$-component of the magnetization of the two
aforementioned models at $|J| \beta = 500$ is
$|Diff (M_z)_{Fe}^{AF}| \lesssim 10^{-436}$ in the interval
of $\frac{h}{|J|} \in [0.98, 1.05]$ and 
$|Diff (M_z)_{Fe}^{AF}| \lesssim 10^{-825}$ in the 
interval  of $\frac{h}{|J|}\in [1.9, 2.1]$. For $|J| \beta =100$
these differences decrease and they are $\leq 10^{-89}$.
Increasing the value of  $\frac{D}{|J|}$ (crystal field 
per unit of $|J|$) it is impossible  to experimentaly measure 
the difference  between the  z-component of the ferro 
($J= -1$) and AF ($J=1$) models.

In order to compare the specific heat and the entropy, both per site, of
ferromagnetic BEG model with $\frac{K}{|J|} = -2$ and the spin-1
AF Ising model, we define the percent  differences
\begin{eqnarray}   \label{4.10}
&&PD (L)_{Fe}^{AF} \left(\frac{h}{|J|}, \frac{D}{|J|}; |J| \beta \right) 
\equiv         \nonumber    \\
&\equiv&
\left( \frac{L_{Fe}^{BEG}  \left(-1, \frac{h}{|J|}, \frac{D}{|J|},
           \frac{K}{|J|} = -2; |J| \beta \right)
  - L_{AF}^{Ising}  \left(1, \frac{h}{|J|}, \frac{D}{|J|},
           \frac{K}{|J|} = 0; |J| \beta \right) }
  {L_{Fe}^{BEG}  \left(-1, \frac{h}{|J|}, \frac{D}{|J|},
           \frac{K}{|J|} = -2; |J| \beta \right)}  
\right) \times 100\%,    \nonumber  \\
\end{eqnarray}

\noindent where $L$ can be the specific heat per site, ${\cal C}$, or
the entropy per site, ${\cal S}$.

First, let us compare the specific heat per site of the
two models. First we consider $\frac{D}{|J|} = 0.51$
at $|J| \beta = 500$. We obtain that 
$|PD (C)_{Fe}^{AF}| \lesssim 10^{-6} \%$ in the interval 
$\frac{h}{|J|} \in [0, 0.5]$, but at $\frac{h}{|J|} =0.51$, 
in which occurs the $A \rightleftharpoons E/E^{\prime}$ 
transition in the diagrams of Fig.\ref{fig_6}, at $T=0$, 
the value of the specific heat per site at each model is
\begin{subequations}

\begin{eqnarray}   
{\cal C}_{Fe}^{BEG} \left( -1, \frac{h}{|J|}= 0.51, \frac{D}{|J|} = 0.51,
     \frac{K}{|J|} = -2; |J| \beta = 500  \right) &\approx & 3.04 \times 10^{-213}
     \label{4.11a}  \\
{\cal C}_{AF}^{Ising} \left(1, \frac{h}{|J|} = 0.51, \frac{D}{|J|} = 0.51,
     \frac{K}{|J|} = 0; |J| \beta = 500  \right) &\approx & 8.70 \times 10^{-8},
        \label{4.11b}
\end{eqnarray}

\end{subequations}

\noindent showing a difference of 205 orders of magnitude in the
specific heat per site of the two models. We do not know which 
mechanism permits so huge difference between these two thermodynamic
functions.

For $\frac{h}{|J|} \gtrsim 0.52$, the percent difference (\ref{4.10})
of the specific heat per site at $|J| \beta = 500$ is
$|PD (C)_{Fe}^{AF}| \lesssim 10^{-8} \%$.

Keeping $\frac{D}{|J|} = 0.51$ and increasing the temperature
up to $|J| \beta =100$, we plot in Fig.\ref{fig_8}a the specific
heat per site  versus $\frac{h}{|J|}$  of the ferromagnetic BEG model
with $\frac{K}{|J|} = -2$ and the spin-1 AF Ising model in the
interval $\frac{h}{|J|} \in [0.4, 0.6]$, showing that the two 
curves do not coincide in this whole interval. For 
$\frac{h}{|J|} \gtrsim 0.8$ we have 
$|PD (C)_{Fe}^{AF}| \lesssim 10^{-12} \%$, and at 
$\frac{h}{|J|} = 1.51$, when we have the transition of phases
$B \rightleftharpoons E/E^{\prime}$ in diagrams \ref{fig_6}
at $T=0$, we obtain 
$PD (C)_{Fe}^{AF} \left(\frac{h}{|J|} = 1.51, \frac{D}{|J|} = 0.51;
|J| \beta = 100  \right) \approx 4.59 \times 10^{-43} \%$.

For $\frac{D}{|J|} =1$ we obtain in the whole interval of
$\frac{h}{|J|}$, that is, $\frac{h}{|J|} \in [0, \infty)$,
\begin{subequations}

\begin{eqnarray}
|PD (C)_{Fe}^{AF} \left(\frac{h}{|J|}, \frac{D}{|J|} = 1;
         |J| \beta = 500  \right)|  & \lesssim &  10^{-216} \%,
               \label{4.12a}  \\
|PD (C)_{Fe}^{AF} \left(\frac{h}{|J|}, \frac{D}{|J|} = 1;
         |J| \beta = 100  \right)|  & \lesssim & 10^{-42} \%.
               \label{4.12b} 
\end{eqnarray}

\end{subequations}

Fig.\ref{fig_9} shows
the percent difference (\ref{4.10}) for the
entropy per site, $L= {\cal S}$, with $\frac{D}{|J|} = 0.51$
at $|J| \beta = 500$. For these values of $\frac{D}{|J|}$ and 
$|J| \beta$, we obtain
\begin{subequations}

\begin{eqnarray}
PD (S)_{Fe}^{AF} \left(\frac{h}{|J|} = 0.8, \frac{D}{|J|} = 0.51;
         |J| \beta = 500  \right)  & \approx & -4.68 \times  10^{-70} \%,
               \label{4.13a}  \\
|PD (S)_{Fe}^{AF} \left(\frac{h}{|J|} \ge 1.51, \frac{D}{|J|} = 0.51;
         |J| \beta = 500  \right)|  & \lesssim & 10^{-439} \%.
                        \label{4.13b} 
\end{eqnarray}

\end{subequations}

The curve $PD (S)_{Fe}^{AF} \times \frac{h}{|J|}$, 
at $|J| \beta =100$ is plotted in Fig.\ref{fig_9}b in 
the interval $\frac{h}{|J|} \in [0, 0.6]$. The value of the 
entropy per site for both models, for some values of 
$\frac{h}{|J|}$ in the same figure, that is
\begin{subequations}

\begin{eqnarray}
{\cal S}_{Fe}^{BEG} \left( -1, \frac{h}{|J|}= 0, \frac{D}{|J|} = 0.51,
     \frac{K}{|J|} = -2; |J| \beta = 100  \right) &\approx & 1.037 \times 10^{-42},
     \label{4.14a}  \\
{\cal S}_{AF}^{Ising} \left( 1, \frac{h}{|J|}= 0, \frac{D}{|J|} = 0.51,
     \frac{K}{|J|} = 0; |J| \beta = 100  \right) &\approx & 1.203 \times 10^{-42},
     \label{4.14b}  
\end{eqnarray}

\noindent and

\begin{eqnarray}
{\cal S}_{Fe}^{BEG} \left( -1, \frac{h}{|J|}= 0.488, \frac{D}{|J|} = 0.51,
     \frac{K}{|J|} = -2; |J| \beta = 100  \right) &\approx & 6.417 \times 10^{-2},
     \label{4.14c}  \\
{\cal S}_{AF}^{Ising} \left( 1, \frac{h}{|J|}= 0.488, \frac{D}{|J|} = 0.51,
     \frac{K}{|J|} = 0; |J| \beta = 100  \right) &\approx & 6.615 \times 10^{-2}.
     \label{4.14d}  
\end{eqnarray}

\end{subequations}

We also obtain that $|PD (S)_{Fe}^{AF} \left(\frac{h}{|J|}\sim 0.7, 
\frac{D}{|J|} = 0.51;|J| \beta = 500  \right) \lesssim 10^{-8}$ and 
$|PD (S)_{Fe}^{AF} \left(\frac{h}{|J|} \gtrsim 1.51, \frac{D}{|J|} = 0.51;
|J| \beta = 100  \right) \lesssim 10^{-85}$.

Finally, for $\frac{D}{|J|} =1$, the percent difference (\ref{4.10})
of the entropy per site of the ferromagnetic BEG model, 
with $\frac{K}{|J|} = -2$, and the spin-1 AF Ising model is 
such  that $|PD (S)_{Fe}^{AF} \left(\frac{h}{|J|}, 
\frac{D}{|J|} = 1;|J| \beta = 500  \right) \lesssim 10^{-215}$,
 and $|PD (S)_{Fe}^{AF} \left(\frac{h}{|J|}, 
\frac{D}{|J|} = 1;|J| \beta = 100  \right) \lesssim 10^{-42}$.
Both inequalities are valid for $\frac{h}{|J|} \ge 0$.


\section{Conclusions} \label{sec_5}

We obtain the exact expression of the Helmholtz free energy (HFE)
of the Blume-Emery-Griffiths (BEG) model in the presence of an 
external longitudinal  magnetic field, for arbitrary values 
of the parameters in the Hamiltonian (\ref{2.2}), valid
for $T > 0$. The addition 
of the term $- K (S_i^z)^2 (S_{i+1}^z)^2$ to the Hamiltonian
of the spin-1 Ising model with single-ion anisotropy term
and external longitudinal magnetic field, enriches the 
phase diagram of the latter at $T=0$\cite{JMMM2014}.
Although our results are valid for the ferromagnetic ($J < 0$) 
and AF ($J> 0$) versions of the BEG model, in the present 
paper we restrict our discussion to the results of 
the ferromagnetic BEG model and take $J= -1$. 
From section \ref{sec_3} on, the parameters
of the Hamiltonian (\ref{2.2}) are scaled in units 
of $|J|$  and the inverse of temperature is measured 
as $|J|\beta$.

The thermodynamics  of the ferromagnetic 
BEG model and of its respective phase diagram at $T=0$, 
is analyzed for two different regimes of the 
$\frac{K}{|J|}$-parameter: $\frac{K}{|J|} < -1$ and 
$\frac{K}{|J|} \ge -1$. In both cases, the value of the 
$z$-component of the magnetization, ${\cal M}_z$,
at low temperature, is determined by the ground states of 
the model presented in Figs.\ref{fig_1}. 
The function ${\cal M}_z(\frac{h}{|J|})$ exhibits 
well-defined plateaux up to $|J| \beta \sim 1$
for $\frac{K}{|J|} \ge -1$; 
however, for for $\frac{K}{|J|} \stackrel{<}{\sim} -1$ 
these  plateaux are lost  for 
$|J| \beta \stackrel{>}{\sim} 10$.

The type of degeneracy of the ground state (exponential
or non-exponential) along the lines that separate  the phases 
in the phase diagram of the ferromagnetic BEG model, at $T=0$,
determines the behavior of the entropy and specific heat,
both per site, of the ferromagnetic model at low temperature.
We show that for $\frac{K}{|J|} \ge -1$ the degeneracy of the ground 
state along the lines that separate the phases in 
Fig.\ref{fig_1}b, at $T=0$, does not diverge in the thermodynamic
limit ($ N \rightarrow \infty$). In this region of values
of $\frac{K}{|J|}$, the specific heat and the 
entropy, both per site, vanishes as $T \rightarrow 0$ in 
the presence of any external longitudinal magnetic field.

The lines that separate the phases in the diagram of 
Fig.\ref{fig_1}a, where we have $\frac{K}{|J|} < -1$ 
and $T=0$, are related to ground states with exponential 
degeneracy. We calculated the degeneracy 
of the ground state along each of these lines and 
obtained their entropy per site at 
$T=0$, and then compared these results
with the ones derived from the HFE of the ferromagnetic 
BEG at very low temperature. We have a strong indication 
that the results agree in the limit of $T \rightarrow 0$.

Finally for $\frac{K}{|J|} = -2$, the phase diagram of the
ferromagnetic BEG model (see Fig.\ref{fig_6}b), at $T=0$, with 
$\frac{D}{|J|} > \frac{1}{2}$  is identical to the phase 
diagram of the spin-1 AF Ising model, with single-ion 
anisotropy term, at $T=0$ (see Fig.\ref{fig_6}a).

We compared the thermodynamics of both models at very 
low temperature ($|J| \beta \sim 500$), when the 
largest contribution to their thermodynamic 
is expected to come from their respective ground 
states. At temperature $|J| \beta = 500$,
the function  ${\cal M}_z (\frac{h}{|J|})$ of both
models coincide by at least 1 part in $10^{14}$. 
Such agreement decreases as the temperature increases.

The ferromagnetic and AF nature of the models 
is not apparent  when we measure ${\cal M}_z$ in the 
interval of $\frac{h}{|J|} \in [0, \infty)$. They are 
distinguished, however, by the specific heat per site and the
entropy per site for $\frac{D}{|J|} \sim 0.51$, when $2 D \sim |J|$, and
for values of the external magnetic field such that $2 h \sim |J|$, even
at low temperatures, $|J \beta = 500$. At this temperature 
and with this value of the parameter $\frac{D}{|J|}$, 
the specific heat per site under $\frac{h}{|J|} = 0.51$  
are valued ${\cal C}_{Fe}^{BEG} \sim 10^{-213}$  and 
${\cal C}_{AF}^{Ising} \sim 10^{-8}$ for the BEG and 
Ising models, respectively: their  values differ by 
205 orders of magnitude. We cannot explain this difference.
When the value of the parameter $\frac{D}{|J|}$ is such 
that $2 D \gg |J|$, the specific heat per site and the 
entropy per site are insensitive, at low temperature, 
if the model is ferromagnetic ($J= -1$) or AF ($J= 1$).

The information about the exact thermodynamics of the staggered BEG 
model in the presence of an external longitudinal magnetic field 
can be obtained from the thermodynamics of the BEG model through 
the relation (\ref{2.4}). The study of the AF BEG model is currently 
under way. We expect to present our results in the near 
future.


\vspace{1cm}

\noindent {\bf Acknowledgements}

M.T. Thomaz thanks Prof. J.F. Stilck for interesting 
discussions about the degeneracy in the phase diagram 
of the BEG model.


\appendix

\setcounter         {equation}{0}
\def\theequation{A.\arabic{equation}}

\section{The exact HFE of the $D=1$ BEG  model in the \\ presence 
of a longitudinal magnetic field.}  \label{Apend_A}

The transfer matrix  method\cite{kramers1,kramers2,baxter} 
was used by Krinsky and Furman\cite{krinsky} to calculate the HFE of the
BEG model in the presence of a longitudinal magnetic 
field ($L=0$ in Hamiltonian (1.1) of Ref.\cite{krinsky}). 

In this appendix we follow the steps of Ref.\cite{JMMM2014},
but here we  write  the root of cubic equation (\ref{A.4}) with the 
largest  modulus as a real number, obtaining the exact HFE 
of the model (\ref{2.2}).

\vspace{0.3cm}

The partition function ${\mathcal Z} (J, h, D; \beta)$ 
of the BEG Hamiltonian (\ref{2.2}) is equal to\cite{baxter}
\begin{eqnarray} \label{A.1}
{\mathcal Z} (J, h, D, K; \beta) = Tr[{\bf U}^N],
\end{eqnarray}
 
\noindent where $N$ is the number of sites in the periodic chain
and the matrix {\bf U} for the symmetric Hamiltonian (\ref{2.2}) is
\begin{eqnarray}  \label{A.2}
{\bf U} (J, h, D, K; \beta) =
\left[
   \begin{array}{c c c}
  e^{- \beta(J + 2h+2D - K)} & e^{- \beta( h + D)} & e^{- \beta( -J + 2D - K)} \\
  e^{- \beta( h + D)} & 1 & e^{- \beta( -h + D)}  \\
  e^{- \beta( -J + 2D - K)} & e^{- \beta( -h + D)} &  e^{- \beta(J - 2h+2D - K)}
   \end{array}
\right]  ,
\end{eqnarray}

\noindent and $\beta = \frac{1}{k T}$, in which $k$ is  the
Boltzmann's constant and $T$ is the absolute temperature 
in kelvin.

The matrix ${\bf U} (J, h, D, K; \beta)$ 
is hermitian for any value of $J$, $h$, $D$, $K$ and $\beta$. 
Its three eigenvalues are  real. The matrices ${\bf U}$ 
[see eq.(\ref{A.2})] and ${\bf T}$  [in Ref.\cite{krinsky}] 
differ by a rearrangement of  lines and the sign of the
external magnetic field $h$. The partition function (\ref{A.1})
is an even function of $h$.

In the thermodynamic limit ($N \rightarrow \infty$), the partition 
function (\ref{A.1}) of the model and its HFE ${\mathcal W}$ 
are, respectively,
\begin{eqnarray}  \label{A.3}
{\mathcal Z} (J, h, D, K; \beta) = (\lambda_1)^N  
   \hspace{0.4cm} \mbox{and} \hspace{0.4cm}
 {\mathcal W} (J, h, D, K; \beta) = - \frac{1}{\beta}
\; \; \ln[\lambda_1 (J, h, D; \beta)], 
\end{eqnarray}

\noindent for non-degenerate eigenvalues of {\bf U}. We assume that
$\lambda_1$ is the eigenvalue of matrix {\bf U} with the largest
modulus, root of the cubic equation 
\begin{eqnarray}  \label{A.4}
- \lambda^3 + P \lambda^2 + Q \lambda + R = 0  ,
\end{eqnarray}

\noindent in which
\begin{subequations}
\begin{eqnarray}   
P  &=& 1 + 2 e^{-\beta (J+2D - K)} \cosh(2 \beta h) = tr[{\bf U}] 
                             \label{A.5a} \\
Q &=& 4 e^{-2 \beta D} \;\;  e^{- \frac{\beta (J-K)}{2}}
\cosh(2 \beta h) \sinh\left(\frac{\beta (J-K)}{2}\right)  
+  2  e^{- 2\beta (2D-K)}  \;\; \sinh(2 \beta J),  
         \nonumber     \\
       &&            \label{A.5b} \\
R &=& -2 e^{-2 \beta(2D - K)} \sinh(2 \beta J)   \;\;
+ 4 e^{-\beta(4D - K)} \; \sinh(\beta J).   \label{A.5c}
\end{eqnarray}
\end{subequations}
The uniqueness of this eigenvalue for the matrix (\ref{A.2}), is ensured
by the Perron-Frobenius Theorem\cite{simon93} for any temperature $T$.

The roots of the cubic equation  (\ref{A.4})  are well
known\cite{schaum}. The root with the largest modulus is
\begin{eqnarray} \label{A.6}  
\lambda_1 = 2  \sqrt{- \tilde{Q}} \;\; \cos\left(\frac{\theta}{3}\right) + \frac{P}{3}  
\end{eqnarray}

\begin{subequations}

\noindent where
\begin{eqnarray}  \label{A.7a}
\cos(\theta)  = \frac{\tilde{R}}{\sqrt{ (- \tilde{Q})^3}}    
\end{eqnarray}

\noindent with
\begin{eqnarray}   \label{A.7b}
\tilde{Q} = - \frac{3 Q + P^2}{9} 
     \hspace{1cm} \rm{and}  \hspace{1cm} 
\tilde{R} = \frac{ 9 Q P + 27 R + 2 P^3}{54}  .
\end{eqnarray}

\end{subequations}

The expression of $\lambda_1$ has cubic roots. Plotting the
thermodynamic functions of the BEG model required numerical evaluation
of that expression. In this work the CAS {\tt Maple} has been used, and
some spurious complex values (probably due to the way the cubic roots are
handled by the system) appeared in the floating point evaluation even
for $700$ significant digits, so some caution had to be taken.
The expression (\ref{A.3}) of the HFE, valid at any temperature $T$
and obtained from eqs.(\ref{A.6})- (\ref{A.7b}), is exact for the 
ferromagnetic ($J<0$) and the AF ($J>0$) BEG models in the 
presence of a longitudinal external magnetic field.


\setcounter         {equation}{0}
\def\theequation{B.\arabic{equation}}

\section{The states and energies of the dimers }  \label{Apend_B}

For spin-1 there are nine possible dimers in neighbouring sites
of the chain. Those states and their corresponding energies, 
obtainable from eq.(\ref{3.2}) are
\begin{eqnarray}
|D^{(A)} \rangle_{i, i+1} \equiv |0\rangle_i \otimes |0\rangle_{i+1}
& \rm{and} & 
  \frac{\varepsilon^{(A)}_{i, i+1}} {|J|} = 0   ,   \label{B.1} \\
|D^{(B)} \rangle_{i, i+1} \equiv |1\rangle_i \otimes |1\rangle_{i+1}
& \rm{and} & 
\frac{\varepsilon^{(B)}_{i, i+1}} {|J|} = -1 + \frac{2D}{|J|} 
     - \frac{2h}{|J|} - \frac{K}{|J|} ,   \label{B.2} \\
|D^{(C)} \rangle_{i, i+1} \equiv |-1\rangle_i \otimes |-1\rangle_{i+1}
& \rm{and} &  
\frac{\varepsilon^{(C)}_{i, i+1}} {|J|} = -1 + \frac{2D}{|J|} 
     + \frac{2h}{|J|} - \frac{K}{|J|} ,   \label{B.3} 
\end{eqnarray}
\begin{subequations}
\begin{eqnarray}
|D^{(E_1)} \rangle_{i, i+1} \equiv |0\rangle_i \otimes |1\rangle_{i+1}
& \rm{and} &  
\frac{\varepsilon^{(E_1)}_{i, i+1}} {|J|} = \frac{D}{|J|} - \frac{h}{|J|} ,   \label{B.4a} \\
|D^{(E_2)} \rangle_{i, i+1} \equiv |1\rangle_i \otimes |0\rangle_{i+1}
& \rm{and} &  
\frac{\varepsilon^{(E_2)}_{i, i+1}} {|J|} = \frac{D}{|J|} - \frac{h}{|J|} ,   \label{B.4b} 
\end{eqnarray}
\end{subequations}
\begin{subequations}
\begin{eqnarray}
|D^{(F_1)} \rangle_{i, i+1} \equiv |0\rangle_i \otimes |-1\rangle_{i+1}
& \rm{and} &  
\frac{\varepsilon^{(F_1)}_{i, i+1}} {|J|} = \frac{D}{|J|} + \frac{h}{|J|} ,   \label{B.5a} \\
|D^{(F_2)} \rangle_{i, i+1} \equiv |-1\rangle_i \otimes |0\rangle_{i+1}
& \rm{and} &  
\frac{\varepsilon^{(F_2)}_{i, i+1}} {|J|} = \frac{D}{|J|} + \frac{h}{|J|} ,   \label{B.5b} 
\end{eqnarray}
\end{subequations}
\begin{subequations}
\begin{eqnarray}
|D^{(G_1)} \rangle_{i, i+1} \equiv |1\rangle_i \otimes |-1\rangle_{i+1}
& \rm{and} &  
\frac{\varepsilon^{(G_1)}_{i, i+1}} {|J|} = 1+ \frac{2D}{|J|} - \frac{K}{|J|} ,   \label{B.6a} \\
|D^{(G_2)} \rangle_{i, i+1} \equiv |-1\rangle_i \otimes |1\rangle_{i+1}
& \rm{and} &  
\frac{\varepsilon^{(G_2)}_{i, i+1}} {|J|} = 1+ \frac{2D}{|J|} - \frac{K}{|J|}  .   \label{B.6b} 
\end{eqnarray}
\end{subequations}


\setcounter         {equation}{0}
\def\theequation{C.\arabic{equation}}

\section{The ground states and energies of the BEG model
in the presence of a longitudinal magnetic field }  \label{Apend_C}

We assume that the chain has a even number of sites $N$ 
by letting $N = 2M$, in which $M$ is a positive integer.
In the thermodynamic limit ($N \rightarrow \infty$), we also have 
$M \rightarrow \infty$. The ground state vectors at each phase in 
the phase diagram in Fig.\ref{fig_1}b at $T=0$ are

\vspace{-0.5cm}

\begin{subequations}

\begin{eqnarray}
|\Psi_0 \rangle_A & =& |0\rangle_1 \otimes  |0\rangle_2 \otimes 
\cdots   \otimes  |0\rangle_{2M},   \label{C.1a} \\
|\Psi_0 \rangle_B & =& |1\rangle_1 \otimes  |1\rangle_2 \otimes 
\cdots   \otimes  |1\rangle_{2M},   \label{C.1b} \\
|\Psi_0 \rangle_C & =& |-1\rangle_1 \otimes  |-1\rangle_2 \otimes 
\cdots   \otimes  |-1\rangle_{2M},   \label{C.1c} \\
|\Psi_0 \rangle_E & =& |0\rangle_1 \otimes  |1\rangle_2 \otimes 
|0\rangle_3 \otimes  |1\rangle_4 \otimes
\cdots   \otimes  |0\rangle_{2M-1}  \otimes |1\rangle_{2M},   \label{C.1d} \\
|\Psi_0 \rangle_{E^{\prime}} & =& |1\rangle_1 \otimes  |0\rangle_2 \otimes 
|1\rangle_3 \otimes  |0\rangle_4 \otimes
\cdots   \otimes  |1\rangle_{2M-1}  \otimes |0\rangle_{2M},   \label{C.1e} \\
|\Psi_0 \rangle_F & =& |0\rangle_1 \otimes  |-1\rangle_2 \otimes 
|0\rangle_3 \otimes  |-1\rangle_4 \otimes
\cdots   \otimes  |0\rangle_{2M-1}  \otimes |-1\rangle_{2M},   \label{C.1f} \\
|\Psi_0 \rangle_{F^{\prime}} & =& |-1\rangle_1 \otimes  |0\rangle_2 \otimes 
|-1\rangle_3 \otimes  |0\rangle_4 \otimes
\cdots   \otimes  |-1\rangle_{2M-1}  \otimes |0\rangle_{2M}.   \label{C.1g} 
\end{eqnarray}

\end{subequations}

\noindent One is reminded that $S_i^z |s\rangle_i = s |s\rangle_i$, 
$s \in \{ 0, \pm 1\}$ and $i \in\{ 1, 2, \cdots 2M\}$. 

The states $|\Psi_0 \rangle_{E^{\prime}}$, $|\Psi_0 \rangle_{F^{\prime}}$
and $|\Psi_0 \rangle_{G^{\prime}}$ have the same energies as the
states $|\Psi_0 \rangle_E$, $|\Psi_0 \rangle_F$ and $|\Psi_0 \rangle_G$,
respectively.

The values of the ground state energy of the phases 
$A$, $B$, $E  (E^{\prime})$ and $F  (F^{\prime})$ for 
$\frac{h}{|J|} \ge 0$, are, respectively,

\vspace{-0.5cm}

\begin{subequations}
\begin{eqnarray}
\frac{E_0^A}{|J|}  & = & 0,   \label{C.2a}  \\
\frac{E_0^B}{|J|}  & = & 2M \left( \frac{J}{|J|} + \frac{2D}{|J|}
       - \frac{2h}{|J|} - \frac{K}{|J|} \right) ,   \label{C.2b}  \\
\frac{E_0^E}{|J|}  & = & 2M \left( \frac{D}{|J|} - \frac{h}{|J|} \right)
   =  \frac{E_0^{E^{\prime}}}{|J|},   \label{C.2c}  \\
\frac{E_0^F}{|J|}  & = & 2M \left( \frac{D}{|J|} + \frac{h}{|J|} \right)
   =  \frac{E_0^{F^{\prime}}}{|J|}.   \label{C.2d}  
\end{eqnarray}

\end{subequations}

The phase diagram of the AF spin-1 Ising model, with a single-ion anisotropy 
term, in the presence of an external longitudinal magnetic field has an
extra phase \cite{JMMM2014} at $T=0$, the N\'eel state, given by
the vector states
\begin{subequations}
\begin{eqnarray}
|\Psi_0 \rangle_G & =& |1\rangle_1 \otimes  |-1\rangle_2 \otimes 
|1\rangle_3 \otimes  |-1\rangle_4 \otimes
\cdots   \otimes  |1\rangle_{2M-1} |-1\rangle_{2M},   \label{C.3a}  \\
|\Psi_0 \rangle_{G^{\prime}} & =& |-1\rangle_1 \otimes  |1\rangle_2 \otimes 
|-1\rangle_3 \otimes  |1\rangle_4 \otimes
\cdots   \otimes  |-1\rangle_{2M-1} |1\rangle_{2M}.   \label{C.3b}
\end{eqnarray}

\end{subequations}

The ground state energy, in units of $|J|$, of theses states $G$  and $
G^{\prime}$ is 
\begin{eqnarray}   \label{C.4}
\frac{E_0^G}{|J|}  = \frac{E_0^{G^{\prime}}}{|J|}
=  2M \left( -\frac{J}{|J|} + \frac{2D}{|J|}  - \frac{K}{|J|}  \right) 
      =  \frac{E_0^{G^{\prime}}}{|J|}.   
\end{eqnarray}


\setcounter         {equation}{0}
\def\theequation{D.\arabic{equation}}

\section{Degeneracy of ground states along transition lines and on
critical points of phase diagrams of the ferromagnetic BEG model
with $\frac{K}{|J|} < -1$ at $T=0$} \label{Apend_D}

Let us take the transition \eq{A \rightleftharpoons E/E^\prime} as an
example for the calculation of the degeneracy of the ground states
along the boundary of two phases in the phase diagram in
Fig.\ref{fig_1}a {{of the ferromagnetic BEG model at $T=0$}}. 
The critical point ${\cal T}$, excluded from 
this analysis, will be treated subsequently.
For all points in the vicinity of (but not upon) this line, 
the chain may be found in either one of the following states, 
among those listed in Appendix \ref{Apend_C}:
\begin{eqnarray}
	|\Psi_0 \rangle_A & =& |0\rangle_1 \otimes  |0\rangle_2 \otimes 
	\cdots   \otimes  |0\rangle_{2M},    \label{D.1}\\
	|\Psi_0 \rangle_E & =& |0\rangle_1 \otimes  |1\rangle_2 \otimes 
	|0\rangle_3 \otimes  |1\rangle_4 \otimes
	\cdots   \otimes  |0\rangle_{2M-1}  \otimes |1\rangle_{2M},  
	                   \label{D.2} \\
	|\Psi_0 \rangle_{E^{\prime}} & =& |1\rangle_1 \otimes  |0\rangle_2 \otimes 
	|1\rangle_3 \otimes  |0\rangle_4 \otimes
	\cdots   \otimes  |1\rangle_{2M-1}  \otimes |0\rangle_{2M}.  \label{D.3}
	\end{eqnarray}

\noindent Hence, on the phase transition line, if the $i$-th site (for
\eq{i\in \{1,2,\dots 2M-1\}}, in which $M$ is an integer) happens to be
in the state \eq{|0\rangle_i}, the state of the $(i+1)$-th site may be
either \eq{|0\rangle_{i+1}} or \eq{|1\rangle_{i+1}}. On the other hand,
if the $i$-th site state is \eq{|1\rangle_i}, the state of the
$(i+1)$-th site must be \eq{|0\rangle_{i+1}}. 
In other words, the energy constraint on this phase transition line 
imposes some {\it sequencing rules} on the states of the chain: each 
\eq{\state 0 {i+1}} succeeds either a \eq{\state 0 i} or a 
\eq{\state 1 i}; and each \eq{\state 1 {i+1}} succeeds a 
\eq{\state 0 i}. (Equivalently: \eq{\state 1 {i+1}} never 
succeeds \eq{\state 1 i}.)

Let \eq{\nstate 0 i} and \eq{\nstate 1 i} be the number of possible
occurences of the one-site states \eq{\state 0 {}} and \eq{\state 1 {}}
at the $i$-th site, respectively. The number of possible occurences for
each state at the $(i+1)$-th site can thus be written as a recurrence
relation of the form
\begin{eqnarray}  \label{D.4}
	g^{(i+1)}=T g^{(i)},
	\end{eqnarray}
	
\noindent in which
\begin{eqnarray}  \label{D.5} 
	g^{(i)} = \left[ \begin{tabular}{c}
		\eq{\nstate 0 {i}} \\
		\ \\
		\eq{\nstate 1 {i}}
		\end{tabular}
		\right],
	\hspace{5mm}
	 g^{(1)} = \left[ \begin{tabular}{c}
		\eq{1} \\
		\ \\
		\eq{1}
		\end{tabular}
		\right],  
	\hspace{5mm}  \rm{and}  \hspace{5mm}
	T = \left[ \begin{tabular}{cc}
		\eq{1} & \eq{1} \\
		\eq{1} & \eq{0}
		\end{tabular}
		\right].
	\end{eqnarray}

\noindent {The configuration $g^{(1)}$ corresponds to the chain
with one site ($i=1$), in which we may have either the state \eq{|0\rangle_{1}} 
or the state \eq{|1\rangle_{1}}.}
	
Such recurrence generalizes to
\begin{eqnarray}   \label{D.6}
	g^{(i+p)}=T^{p} g^{(i)};
	\end{eqnarray}
	
\noindent and hence  we may write, relating the 
\eq{1}-st and \eq{(2M)}-th (last) sites,
\begin{eqnarray}   \label{D.7}
	g^{(2M)}=T^{2M-1} g^{(1)}.
	\end{eqnarray}
The total number of states with the same ground
state energy corresponds to the sum of all possibilities
for the \eq{\state 0 {}} and \eq{\state 1 {}} 
states at the $(2M)$-th state. It is then 
equivalent to the $L^1$-norm of the vector \eq{g^{(2M)}}, 
\begin{eqnarray}   \label{D.8}
	\Omega_{A \rightleftharpoons E/E^\prime} = |g^{(2M)}|_1
	   = \nstate 0 {2M} + \nstate 1 {2M};
	\end{eqnarray}

\noindent hence, we should turn our attention to the 
evaluation of the matrix power \eq{T^{2M-1}} in (\ref{D.7}).
The matrix $T$ can be easily diagonalized, yielding
\begin{eqnarray}   \label{D.9}
	\Lambda_{A \rightleftharpoons E/E^\prime} = \left[ \begin{tabular}{cc}
		\eq{\frac{1}{2}(1+\sqrt{5})} & \eq{0} \\
		\eq{0} & \eq{\frac{1}{2}(1-\sqrt{5})}
		\end{tabular}
		\right]
	\end{eqnarray}
	
\noindent and the corresponding matrix of eigenvectors,
\begin{subequations}
\begin{eqnarray}  \label{D.10a}
	{\mathbf D}_{A \rightleftharpoons E/E^\prime} = \left[ \begin{tabular}{cc}
		\eq{\frac{2}{1+\sqrt{5}}} & \eq{\frac{2}{1-\sqrt{5}}} \\
		\ \\
		\eq{1} & \eq{1}
		\end{tabular}
		\right],
	\end{eqnarray}
	
\noindent so that 
\begin{eqnarray}   \label{D.10b}
	{\mathbf D}_{A \rightleftharpoons E/E^\prime}^{-1}\; T\;
	      {\mathbf D}_{A \rightleftharpoons E/E^\prime} 
	            = \Lambda_{A \rightleftharpoons E/E^\prime}. 
	\end{eqnarray}

\end{subequations}
	
\noindent We rewrite  (\ref{D.7}) as
\begin{eqnarray}   \label{D.11}
	g^{(2M)}= {\mathbf D}_{A \rightleftharpoons E/E^\prime}^{-1}  \;
            \Lambda_{A \rightleftharpoons E/E^\prime}^{N-1} \;
               {\mathbf D}_{A \rightleftharpoons E/E^\prime}\gamma^{(1)}.
	\end{eqnarray}
	
\noindent After some algebra, (\ref{D.8}) yields the degeneracy 
{{of the ground states along the transition line 
$A \rightleftharpoons E/E^\prime$ in Fig.\ref{fig_1}a, at $T=0$,}}
\begin{eqnarray}  \label{D.12}
	\Omega_{A \rightleftharpoons E/E^\prime} = 
	\frac{1}{2^{2M}}
	\left[
		\frac{5 + 3 \sqrt{5}}{10} \left(1 + \sqrt 5\right)^{2M} 
		+ 
		\frac{5 - 3 \sqrt{5}}{10} \left(1 - \sqrt 5\right)^{2M} 
		\right].
	\end{eqnarray}
	
\noindent Thus the corresponding entropy for this transition is
\begin{subequations}
\begin{eqnarray}
	S_{A \rightleftharpoons E/E^\prime} = \lim_{M\rightarrow\infty} 
	\frac{\ln \Omega_{A \rightleftharpoons E/E^\prime}}{2M} 
	&=&  \ln\left(\frac{1 + \sqrt 5}{2}\right) = \ln \varphi,
	             \label{13a} \\
     	&\approx & 0.48121,    \label{D.13b}
	\end{eqnarray}
\end{subequations}

\noindent in which \eq{\varphi} is the celebrated golden ratio.

The reader should notice that the periodic spatial boundary
condition on the chain (cf. section \ref{sec_1})
{\it has not been used at all} in the calculation 
of the degeneracy (\ref{D.12}) and hence on the 
determination of the entropy per site (\ref{D.13b}). 
We shall describe in what follows how  the degeneracy can be 
calculated taking that condition into account; the value of 
the entropy per site (\ref{D.13b}), however, will not change.
By identifying the $(N+1)$-th site of the chain with its 
$1$-st site, the sequencing rules described in the beginning 
of this appendix should also apply to the $N$-th and $1$-st site.
Here we have $N = 2M$. 
Namely, if the chain has \eq{\state 1 {1}} in its $1$-st site,
it cannot have \eq{\state 1 {2M}} in its $2M$-th site.  On the 
other hand, if the chain begins with \eq{\state 0 {1}}, there 
are no restrictions: it can end in either \eq{\state 0 {2M}} or 
\eq{\state 1 {2M}}. The two branches of possibilities --- chain 
starting with \eq{\state 0 {1}} or chain starting with 
\eq{\state 1 {1}} --- can be described separately by the 
$1$-st site conditions
	\begin{eqnarray}   \label{D.14}
		 g^{(1),0} = \left[ \begin{tabular}{c}
			\eq{1} \\
			\ \\
			\eq{0}
			\end{tabular}
			\right],
		\hspace{5mm}
		 g^{(1),1} = \left[ \begin{tabular}{c}
			\eq{0} \\
			\ \\
			\eq{1}
			\end{tabular}
			\right].
		\end{eqnarray}
		
\noindent The version of (\ref{D.7}) upon boundary conditions reads
	\begin{eqnarray}  \label{D.15} 
		g^{(2M)}_{B.C.}= R_0 \; T^{2M-1}\; g^{(1),0} + R_1 \; T^{2M-1}\; g^{(1),1},
		\end{eqnarray}

\noindent in which \eq{\{R_0, R_1\}} are matrices that implement the 
pertinent restrictions to each $1$-st site condition. In the 
	present situation, these matrices have the form
	\begin{eqnarray}   \label{D.16}
		R_0 = \left[ \begin{tabular}{cc}
			\eq{1} & \eq{0} \\
			\eq{0} & \eq{1}
			\end{tabular}
			\right]
		\hspace{5mm} {\rm and} \hspace{5mm}
		R_1 = \left[ \begin{tabular}{cc}
			\eq{1} & \eq{0} \\
			\eq{0} & \eq{0}
			\end{tabular}
			\right].
		\end{eqnarray}
		
\noindent The first term in (\ref{D.15}) relates to the chain 
states with a \eq{\state 0 1} state; \eq{R_0} is simply the 
identity matrix, and there are no restrictions to the which 
state the \eq{2M}-th site may have. On the other hand, the 
second term in (\ref{D.15}) relates to the chain states with 
a \eq{\state 1 1} state; the effect of \eq{R_1} on 
\eq{T^{2M-1}\; g^{(1),1}} is that of discarding all possible 
chain states with a \eq{\state 1 {2M}} state, which would violate 
the	energy condition/sequencing rules. (Notice that, without 
boundary conditions, we would have \eq{R_0} and \eq{R_1} both 
equal to the identity matrix, and (\ref{D.15}) would be reduced 
to (\ref{D.7}). The $1$-st site condition Eq.(\ref{D.5}) is 
equivalent to letting \eq{g^{(1)}=g^{(1),0} + g^{(1),1}}.)
The total number of states with the same ground state energy 
corresponds to
\begin{eqnarray}
	  \label{D.17}   
		\Omega^{B.C.}_{A \rightleftharpoons E/E^\prime} 
		= |g^{(2M)}_{B.C.}|_1 =  g^{(2M)}_{B.C.,\state 0 {}} 
		     +  g^{(2M)}_{B.C.,\state 1 {}},
\end{eqnarray}
		
\noindent analogous to (\ref{D.8}); here, 
\eq{g^{(2M)}_{B.C.,\state 0 {}}} and 
\eq{g^{(2M)}_{B.C.,\state 1 {}}} are the components of
\eq{g^{(2M)}_{B.C.}}. We obtain
	\begin{eqnarray}  \label{D.18}
		\Omega^{B.C.}_{A \rightleftharpoons E/E^\prime} = 
		\frac{1}{2^{2M}}\left[ \left( 1 + \sqrt{5} \right)^{2M} + \left( 1 - \sqrt{5} \right)^{2M}\right],
		\end{eqnarray}
		
\noindent which differs from the degeneracy (\ref{D.12}); however, 
the same entropy per site is obtained from (\ref{D.18}) as it is 
obtained from (\ref{D.12}),
	\begin{eqnarray}    \label{D.19}
		S^{B.C.}_{A \rightleftharpoons E/E^\prime} = 
 \lim_{M\rightarrow\infty} \frac{\ln \Omega^{B.C.}_{A \rightleftharpoons E/E^\prime}}{2M} = 
		S_{A \rightleftharpoons E/E^\prime},
		\end{eqnarray}
		
\noindent given by (\ref{13a}).

\vspace{0.3cm}

The analysis for the critical point ${\cal T}$ is 
analogous. In the vicinity of (but not on) ${\cal T}$, the available
states are
\begin{subequations}
\begin{eqnarray}
	|\Psi_0 \rangle_A & =& |0\rangle_1 \otimes  |0\rangle_2 \otimes 
	\cdots   \otimes  |0\rangle_{2M}, \label{D.20a} \\
	|\Psi_0 \rangle_E & =& |0\rangle_1 \otimes  |1\rangle_2 \otimes 
	|0\rangle_3 \otimes  |1\rangle_4 \otimes
	\cdots   \otimes  |0\rangle_{2M-1}  \otimes |1\rangle_{2M},  
	         \label{D.20b}  \\
	|\Psi_0 \rangle_{E^{\prime}} & =& |1\rangle_1 \otimes  |0\rangle_2 \otimes 
	|1\rangle_3 \otimes  |0\rangle_4 \otimes
	\cdots   \otimes  |1\rangle_{2M-1}  \otimes |0\rangle_{2M}, 
	         \label{D.20c}  \\
	|\Psi_0 \rangle_F & =& |0\rangle_1 \otimes  |-1\rangle_2 \otimes 
	|0\rangle_3 \otimes  |-1\rangle_4 \otimes
	          \label{D.20d} \\
	|\Psi_0 \rangle_{F^{\prime}} & =& |-1\rangle_1 \otimes  |0\rangle_2 \otimes 
	|-1\rangle_3 \otimes  |0\rangle_4 \otimes
	\cdots   \otimes  |-1\rangle_{2M-1}  \otimes |0\rangle_{2M}.
	          \label{D.20e}
	\end{eqnarray}
\end{subequations}

\noindent Hence, for the state ${\cal T}$, the $i$-th site may
 be in any one of the states \eq{\left\{
\state {-1} i, \state 0 i, \state 1 i\right\}}, \eq{i\in
\left\{1,2,\dots, 2M \right\}}. Moreover, for \eq{i\in \left\{1,2,\dots,
2M-1 \right\}}, the state \eq{\state {-1}{i+1}} can only be 
preceeded by \eq{\state 0 i}, the state \eq{\state {0}{i+1}} 
can be preceeded by \eq{\state{-1} i} or
\eq{\state{0} i} or \eq{\state{1} i}, and the state 
\eq{\state {1}{i+1}} can only be preceeded by \eq{\state{0} i}
{{in order to guarantee that the dimer has the least possible 
value of energy}}. (Equivalently: \eq{\state{-1}{i+1}} never 
succeeds either \eq{\state{-1}{i}} or \eq{\state{1}{i}};
\eq{\state{1}{i+1}} never succeeds either \eq{\state{-1}{i}} 
or \eq{\state{1}{i}}.) The recurrence relation  among the number of 
possibilities for each state at a site can be expressed as a 
recurrence relation of the same form as (\ref{D.7}), but now with
\begin{eqnarray}    \label{D.21}
	g^{(i)} = \left[ \begin{tabular}{c}
		\eq{\nstate{-1}{i}} \\
		\ \\
		\eq{\nstate 0 {i}} \\
		\ \\
		\eq{\nstate 1 {i}}
		\end{tabular}
		\right],
	\hspace{5mm}
	g^{(1)} = \left[ \begin{tabular}{c}
		\eq{1} \\
		\ \\
		\eq{1} \\
		\ \\
		\eq{1}
		\end{tabular}
		\right],
	\hspace{5mm} \rm{and} \hspace{5mm}
	T = \left[ \begin{tabular}{ccc}
		\eq{0} & \eq{1} & \eq{0} \\
		\eq{1} & \eq{1} & \eq{1} \\
		\eq{0} & \eq{1} & \eq{0}
		\end{tabular}
		\right].
	\end{eqnarray}

The diagonalization of $T$ leads us to
\begin{subequations}
\begin{eqnarray}    \label{D.22a}  
	\Lambda_{\cal T} = \left[ \begin{tabular}{ccc}
		\eq{2} & \eq{0} & \eq{0} \\
		\eq{0} & \eq{-1} & \eq{0} \\
		\eq{0} & \eq{0} & \eq{0}
		\end{tabular}
		\right]
	\end{eqnarray}

\noindent and
\begin{eqnarray}    \label{D.22b}
	{\mathbf D}_{\cal T}  = \left[ \begin{tabular}{ccc}
		\eq{1} & \eq{1} & \eq{-1} \\
		\eq{2} & \eq{-1} & \eq{0} \\
		\eq{1} & \eq{1} & \eq{1}
		\end{tabular}
		\right].
	\end{eqnarray}
\end{subequations}

\noindent By calculating \eq{g^{(2M)}} as in the previous situation,
{{[see eq.(\ref{D.12})]}}, we obtain the degeneracy at the 
critical point ${\cal T}$ in Fig.\ref{fig_1}a, 
\begin{eqnarray}  \label{D.23}           
	\Omega_{J} = \frac{1}{3}\left(2^{2M+2} -1 \right),
	\end{eqnarray}
	
\noindent  and the corresponding entropy
\begin{subequations}
\begin{eqnarray}
	S_{J} = \lim_{M\rightarrow\infty} \frac{\ln \Omega_{J}}{2M} &=&  \ln(2),
	           \label{D.24a}   \\
      	&\approx & 0.69315.   \label{D.24b}
	\end{eqnarray}
\end{subequations}

Once again, the calculation of the degeneracy under the 
boundary conditions may be carried out in the same fashion 
as for the \eq{{A \rightleftharpoons E/E^\prime}} transition. 
Even though the degeneracy itself differs from that of the 
unconditioned case, 
	\begin{eqnarray}  \label{D.25}
		\Omega_{J}^{B.C.} = 2^{2M} + 1,
		\end{eqnarray}
		
\noindent the entropy per site obtained is the same,
	\begin{eqnarray}    \label{D.26}
	S_{J}^{B.C.} = \lim_{M\rightarrow\infty} \frac{\ln \Omega^{B.C.}_{J}}{2M} = S_{J}.
		\end{eqnarray}

\vspace{0.5cm}

The degeneracies and entropies for the other phase transition lines in
the phase diagram of the ferromagnetic BEG model with $\frac{K}{|J|} <
-1$, at $T=0$, and the critical point ${\cal R}$ can be obtained in the
same fashion.

It is important to point out that in order to calculate 
the entropy per site of the ferromagnetic BEG model along 
phase boundaries and multicritical points at $T=0$ we
can calculate the total number of degenerate ground 
states in the chain without 
taking into account the periodic spatial boundary condition.



\setcounter {figure}{0}


\begin{figure} 
\begin{center}
\includegraphics[scale=0.4,angle= -90]{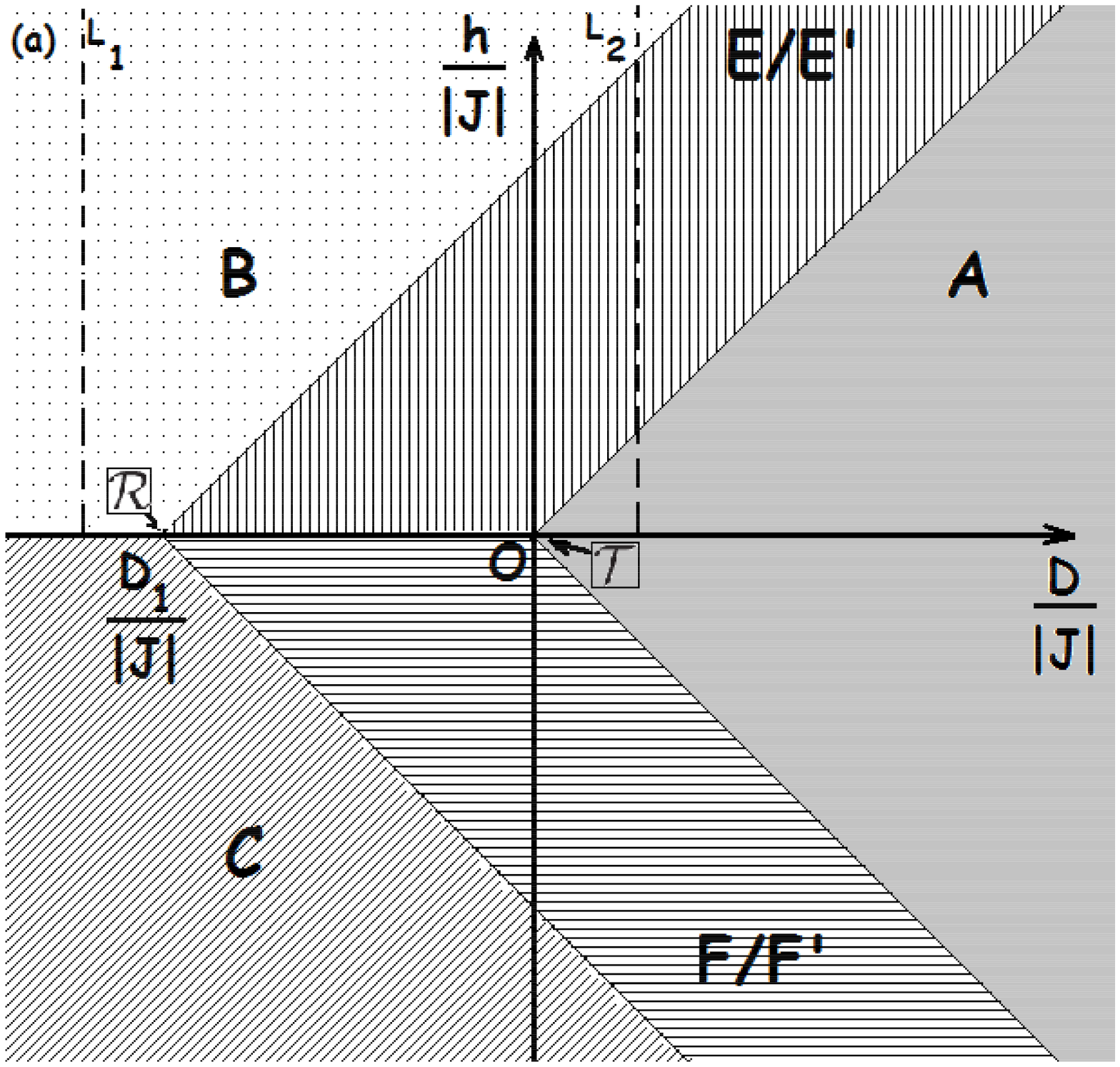}
\hspace{0.5cm}
\hspace{0.5cm}
\includegraphics[scale=0.38,angle= -90]{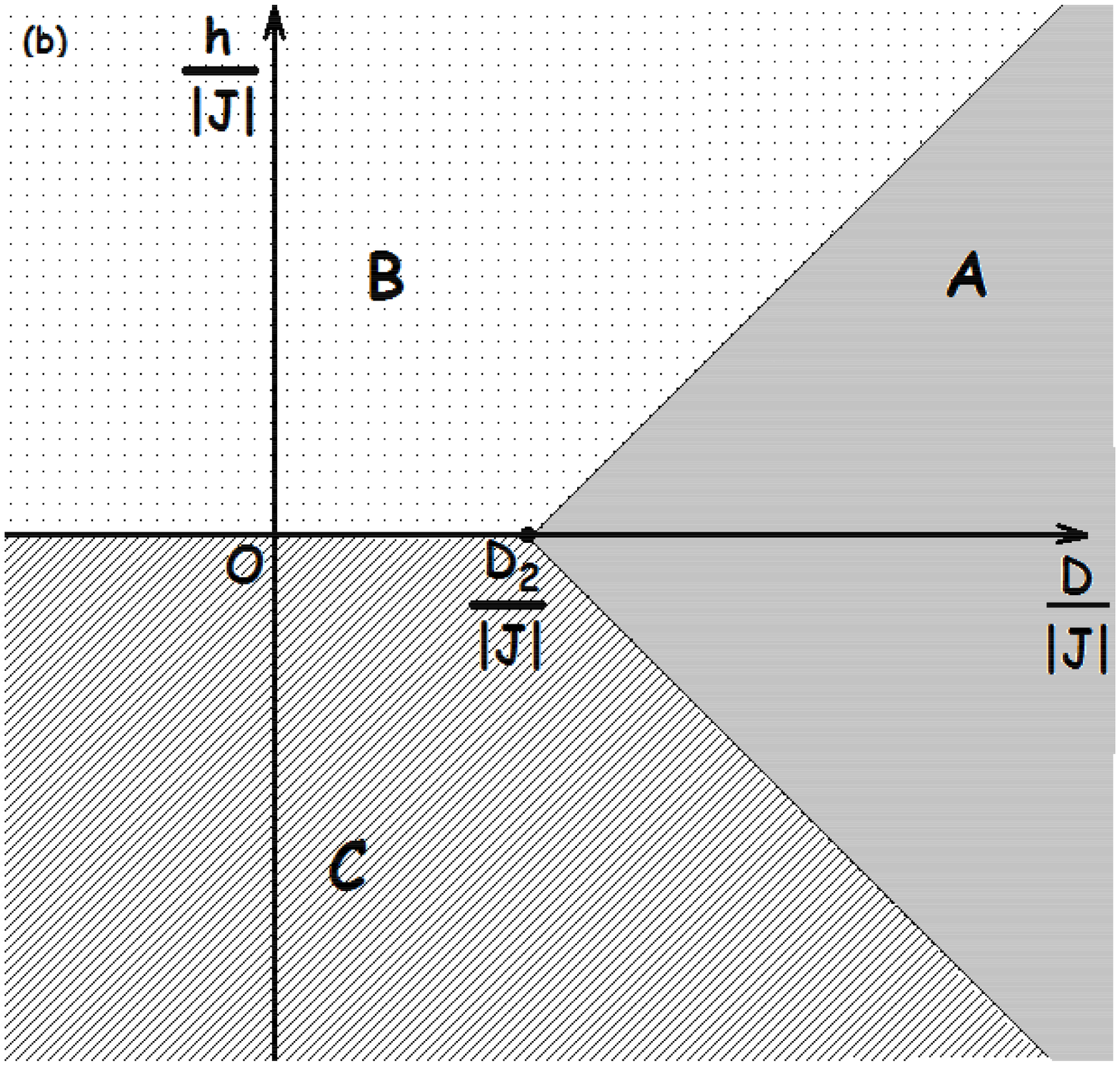}
\vspace*{12mm}
\caption{The phase diagrams, at $T=0$, of the ferromagnetic BEG model
in the presence of an external magnetic field. In $(a)$ we have
$\frac{K}{|J|} < -1$. The vector states corresponding to phases
$A$, $B, \cdots,$ $F/F^{\prime}$ are given in appendix \ref{Apend_C}.
The multicritical point ${\cal R}$ is at  
$\frac{D_1}{|J|} =  1 + \frac{K}{|J|}$. In the phase diagram 
$(b)$ we have $\frac{K}{|J|} \ge -1$. The phases $A$, $B$ and
$C$ in this diagram are the same the ones that appear in 
diagram $(a)$. The tricritical point in this diagram is at 
$\frac{D_2}{|J|} = \frac{1}{2} +  \frac{K}{2|J|}$.
}  \label{fig_1}
\end{center}  
\end{figure}


\begin{figure}
\begin{center}
\includegraphics[scale= 0.35]{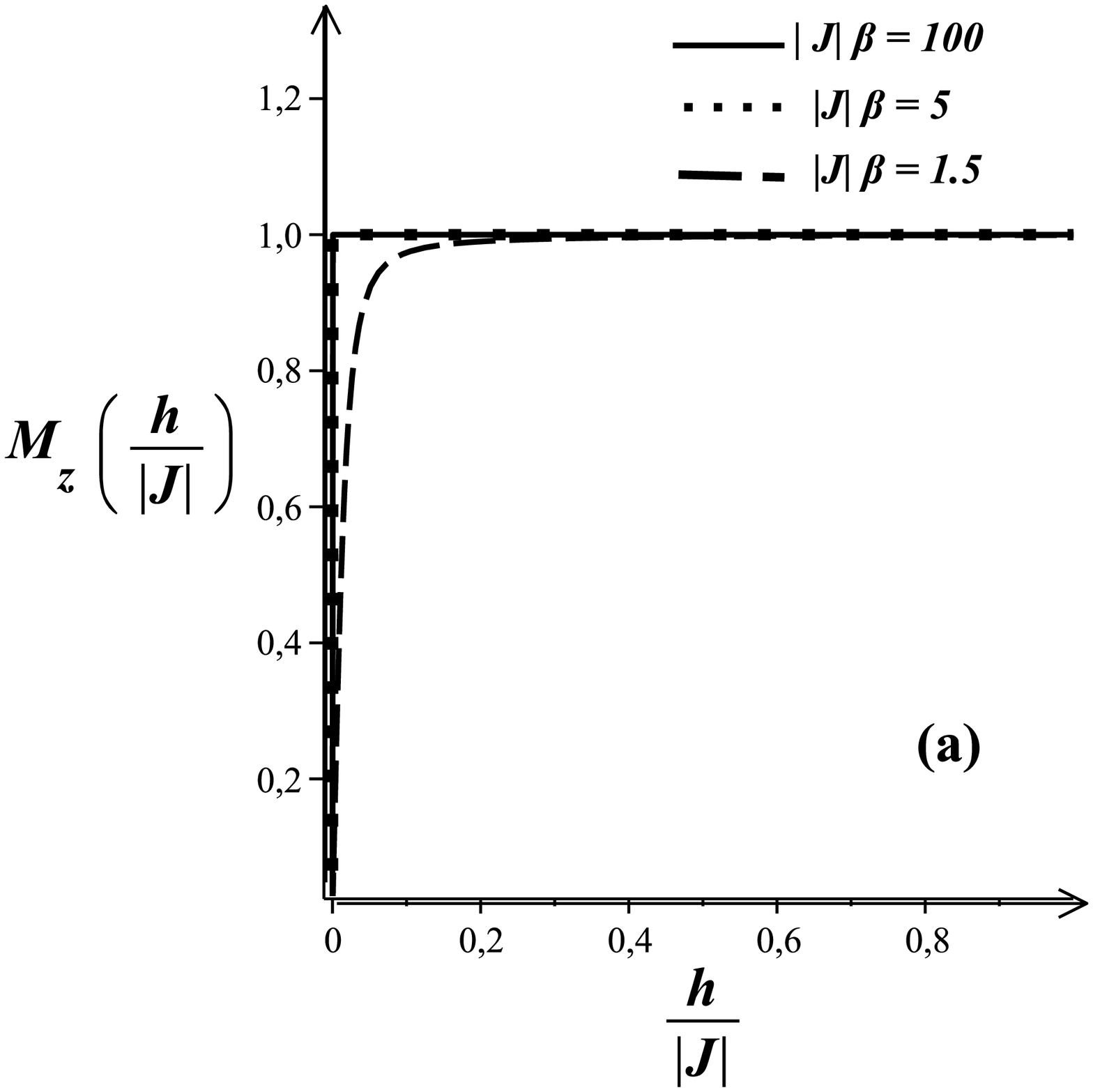} 
\hspace{0.3cm}
\includegraphics[scale= 0.35]{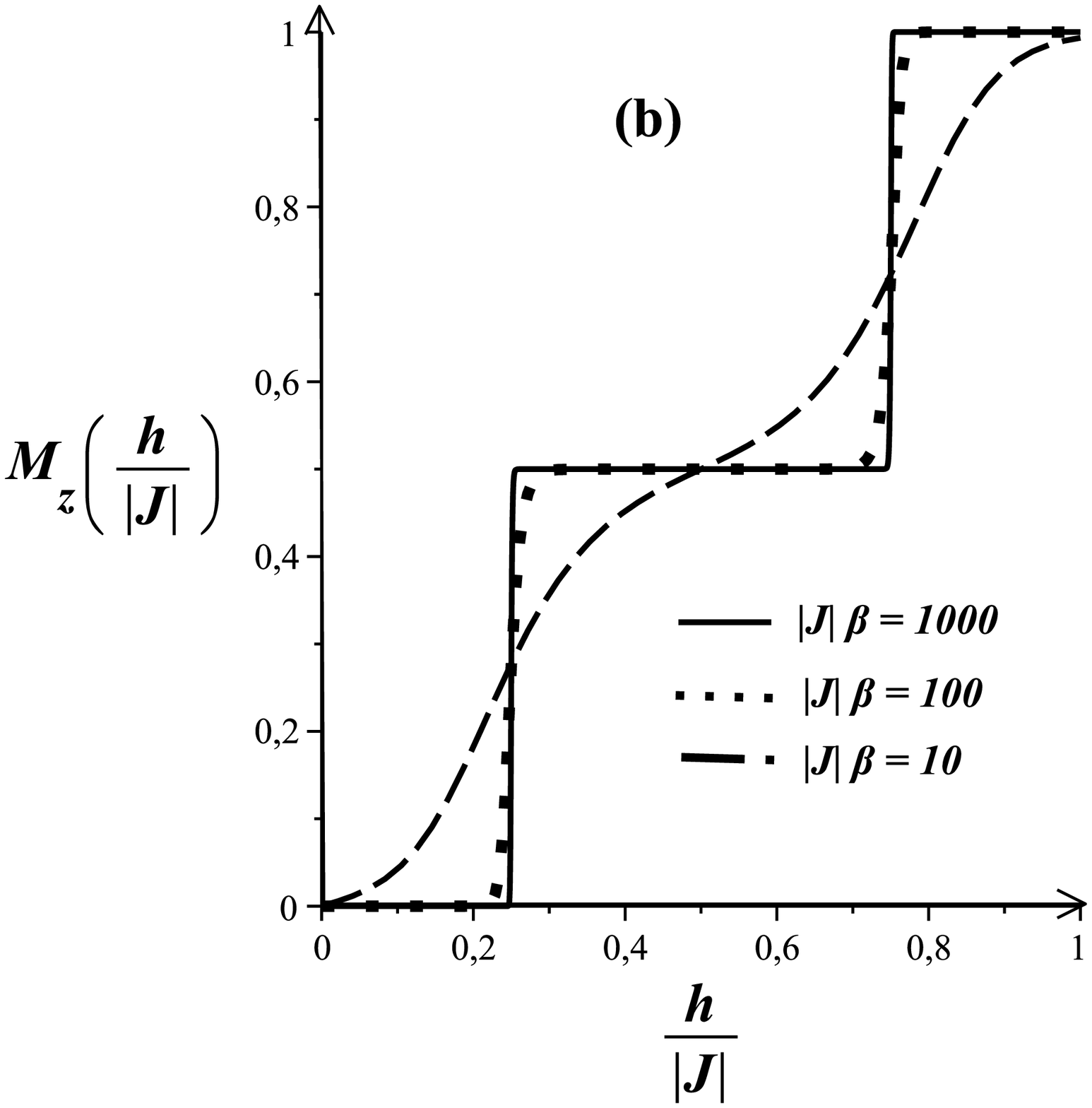} 
\vspace{-0.5cm}
\caption{The z-component of the magnetization ${\cal M}_z$ 
per site as a function of the external magnetic field 
$\frac{h}{|J|}$ for the ferromagnetic BEG model. 
In $(a)$, ${\cal M}_z$ is shown for $\frac{D}{|J|} = - 2$ 
and several values  of $|J| \beta$: 100 (solid line), 
5 (dotted line) and 1.5 (dashed line). In $(b)$  we have 
$\frac{D}{|J|} = 0.25$ with the curves plotted 
for distinct values of $|J| \beta$: 1000 (solid line), 
100 (dotted line) and 10 (dashed line). In both figures we 
have $J= -1$ and $\frac{K}{|J|} = - 1.5$.
   } 
	    \label{fig_2} 
	    
\end{center}
\end{figure}


\begin{figure} 
\begin{center}
\includegraphics[scale= 0.35]{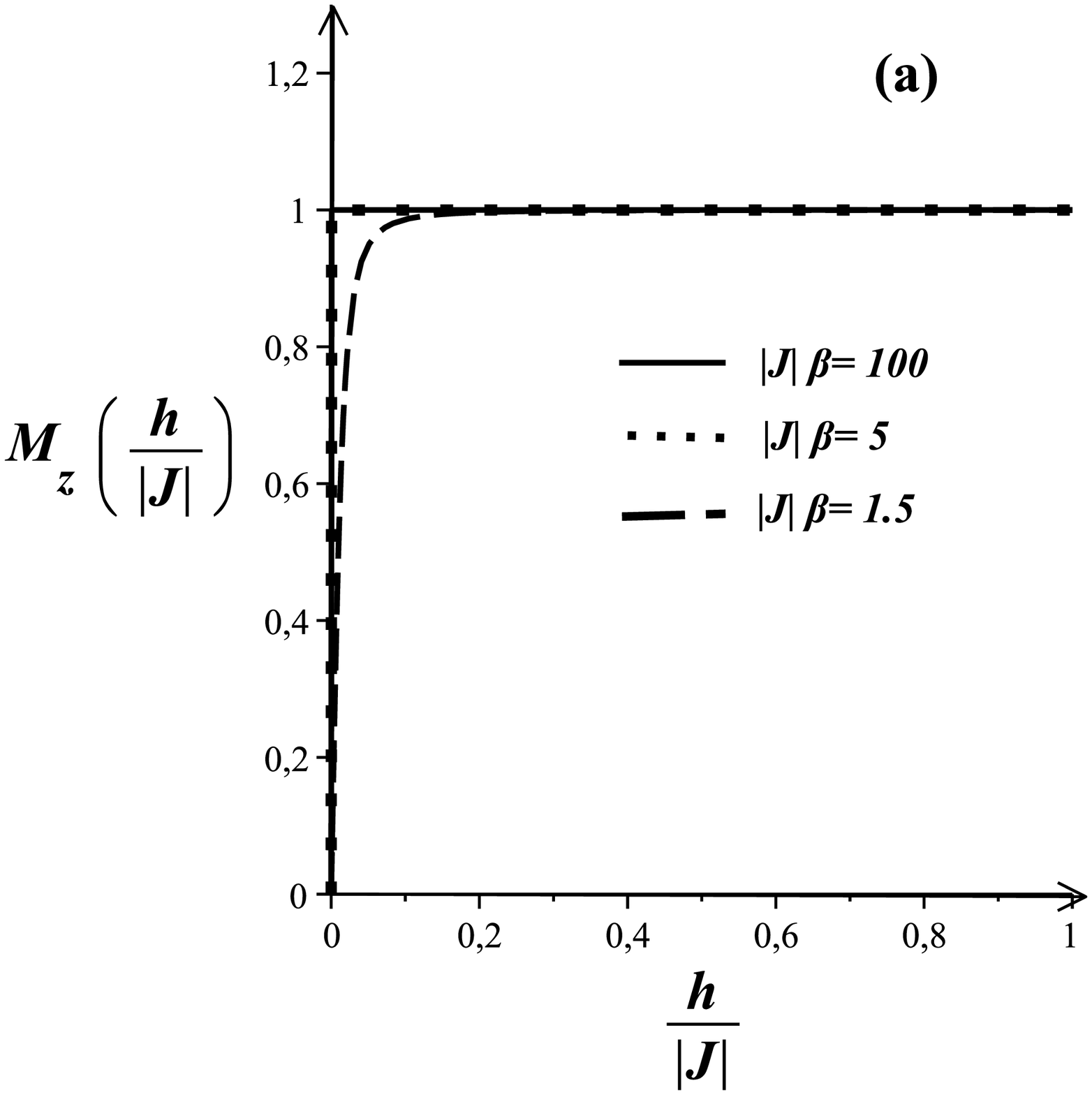} 
\hspace{0.3cm}
\includegraphics[scale= 0.35]{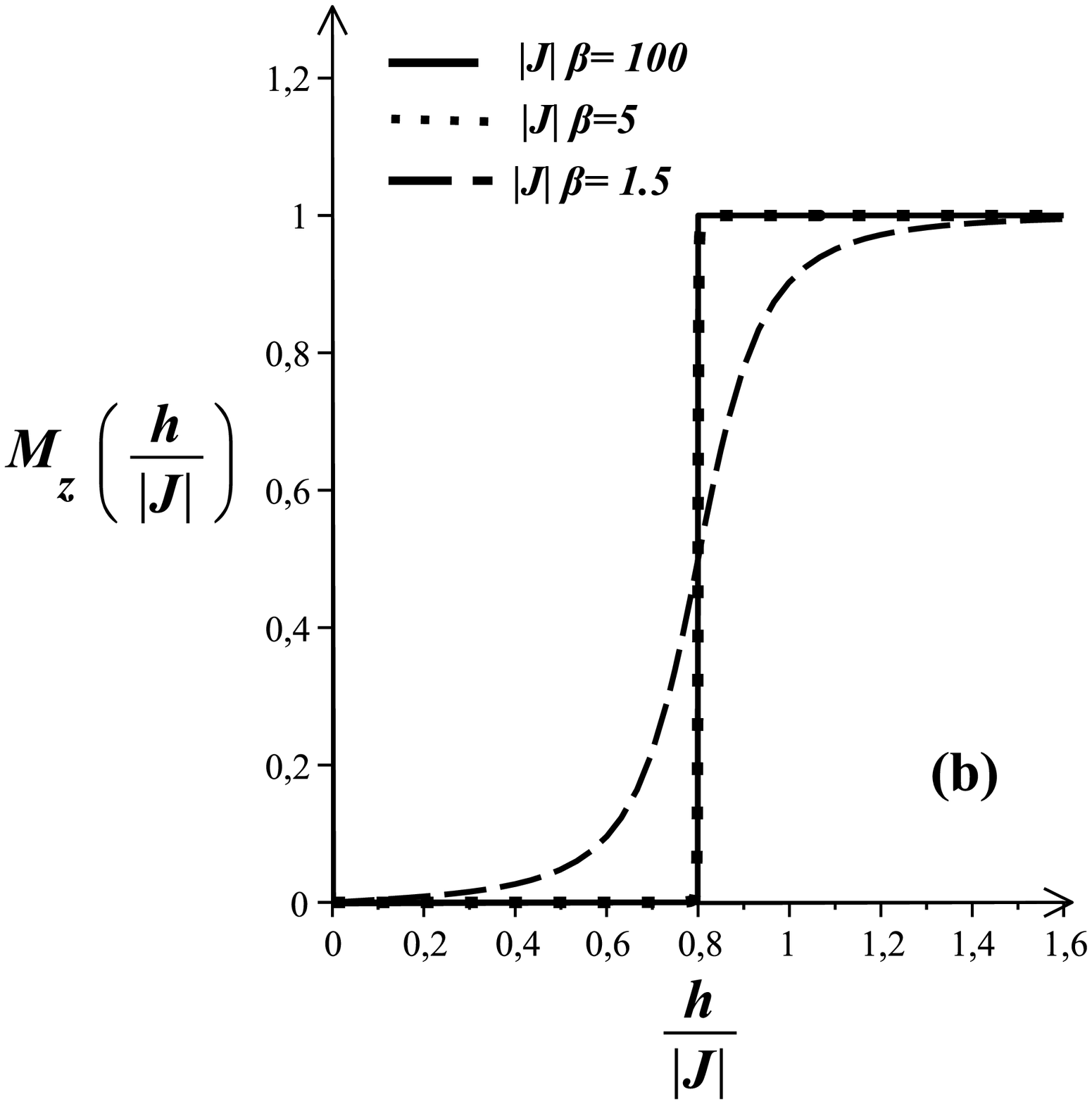}  
\end{center}
\vspace{-1cm}
\caption{ The z-component of the magnetization ${\cal M}_z$ 
per site as a function of the external magnetic field 
$\frac{h}{|J|}$ for the ferromagnetic BEG model in the 
region $\frac{K}{|J|} \ge -1$. In $(a)$ we take 
$\frac{D}{|J|} = - 0.5$. The curves are plotted for 
different values of $|J| \beta$: 100 (solid line), 5 
(dotted line) and 1.5 (dashed line). In $(b)$
we have $\frac{D}{|J|} = 1.8$ and ${\cal M}_z$ is  
plotted  for three values of $|J| \beta$: 
1000 (solid line), 100 (dotted  line) and 10 (dashed line). 
In $(a)$ and $(b)$  we take $J= -1$ and 
$\frac{K}{|J|} = 1$.
}  \label{fig_3}  
 \end{figure}


\begin{figure} 
\begin{center}
\includegraphics[scale= 0.35]{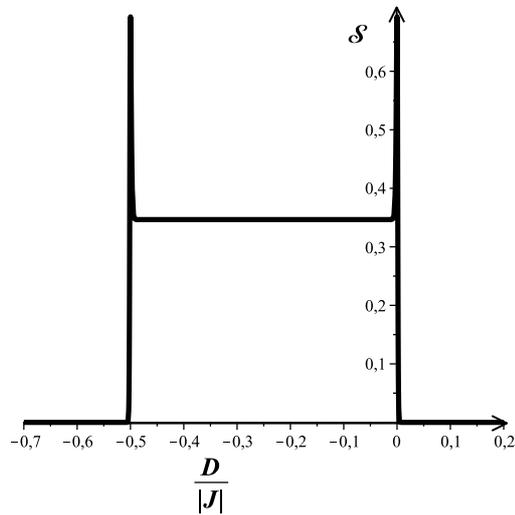}
\hspace{0.1cm}
\end{center}
\vspace{-0.7cm}
\caption{ The entropy per site ${\cal S}$ as a function 
of the single-ion parameter $\frac{D}{|J|}$ is shown
at $|J| \beta = 10^3$, with $\frac{K}{|J|} = -1.5$ and $\frac{h}{|J|} = 0$.
We vary the parameter $\frac{D}{|J|}$ along the horizontal 
line $\frac{h}{|J|} = 0$ in phase diagram of Fig.\ref{fig_1}a 
in order to include in the curve the critical points, at 
$T=0$, ${\cal R}$ ($\frac{D}{|J|} = -0.5$) and 
${\cal T}$ ($\frac{D}{|J|} = 0$).
}   \label{fig_4}   
\end{figure}


\begin{figure} 
\begin{center}
\includegraphics[width=3.5cm,height= 5.5cm,angle= 0]{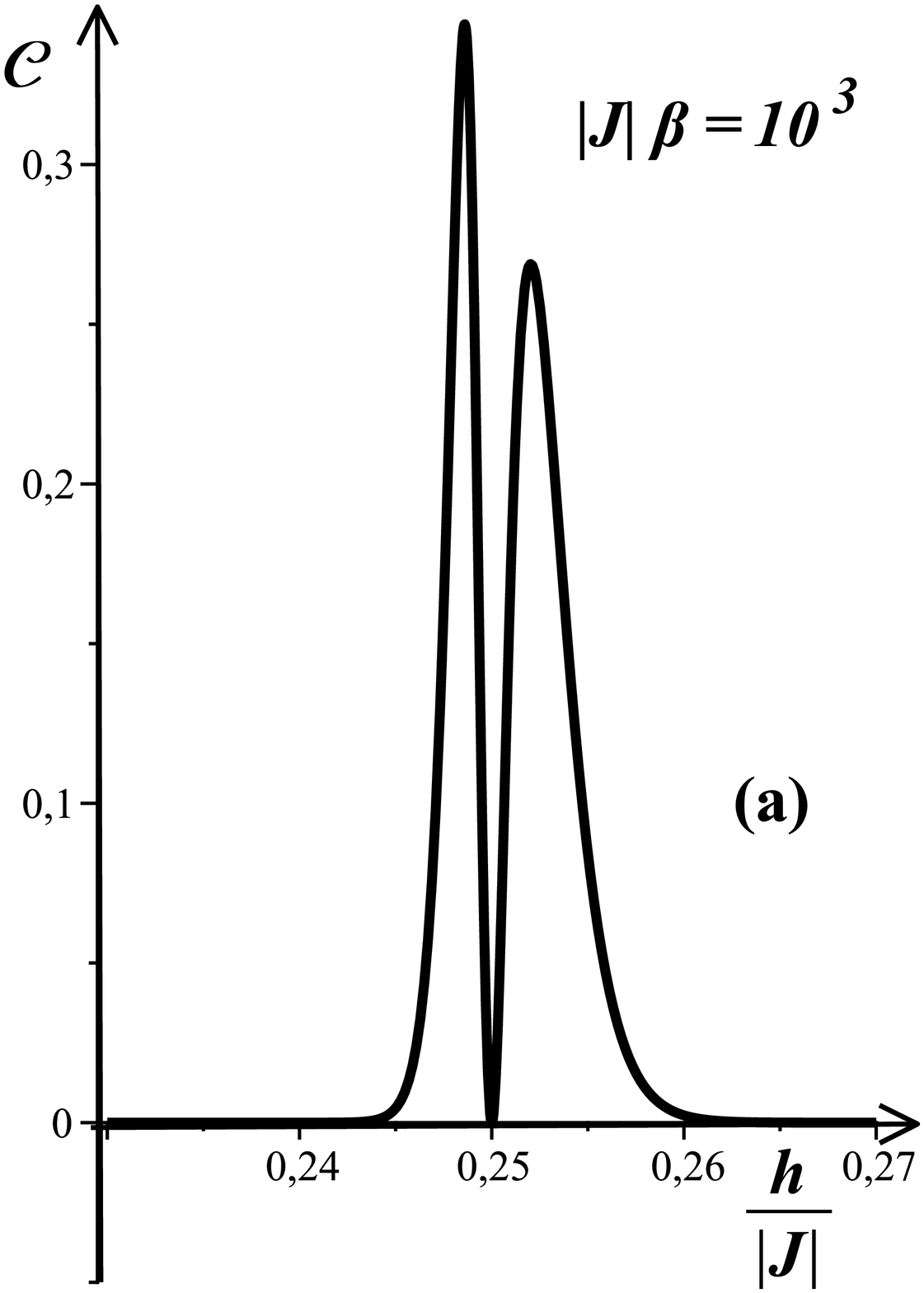} 
\includegraphics[width =3.5cm,height= 5.5cm,angle= 0]{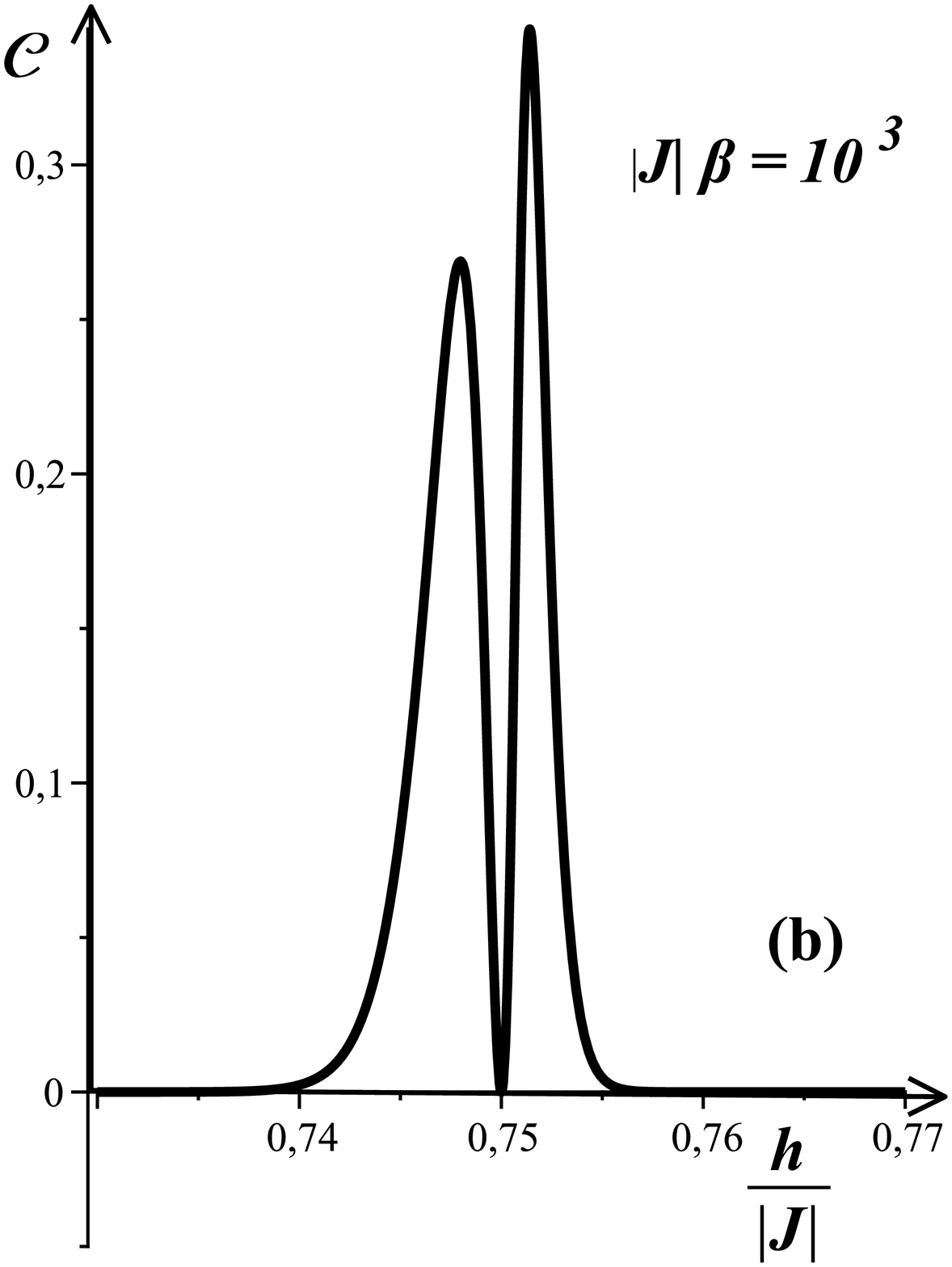} 
\hspace{0.3cm}
\includegraphics[scale= 0.3]{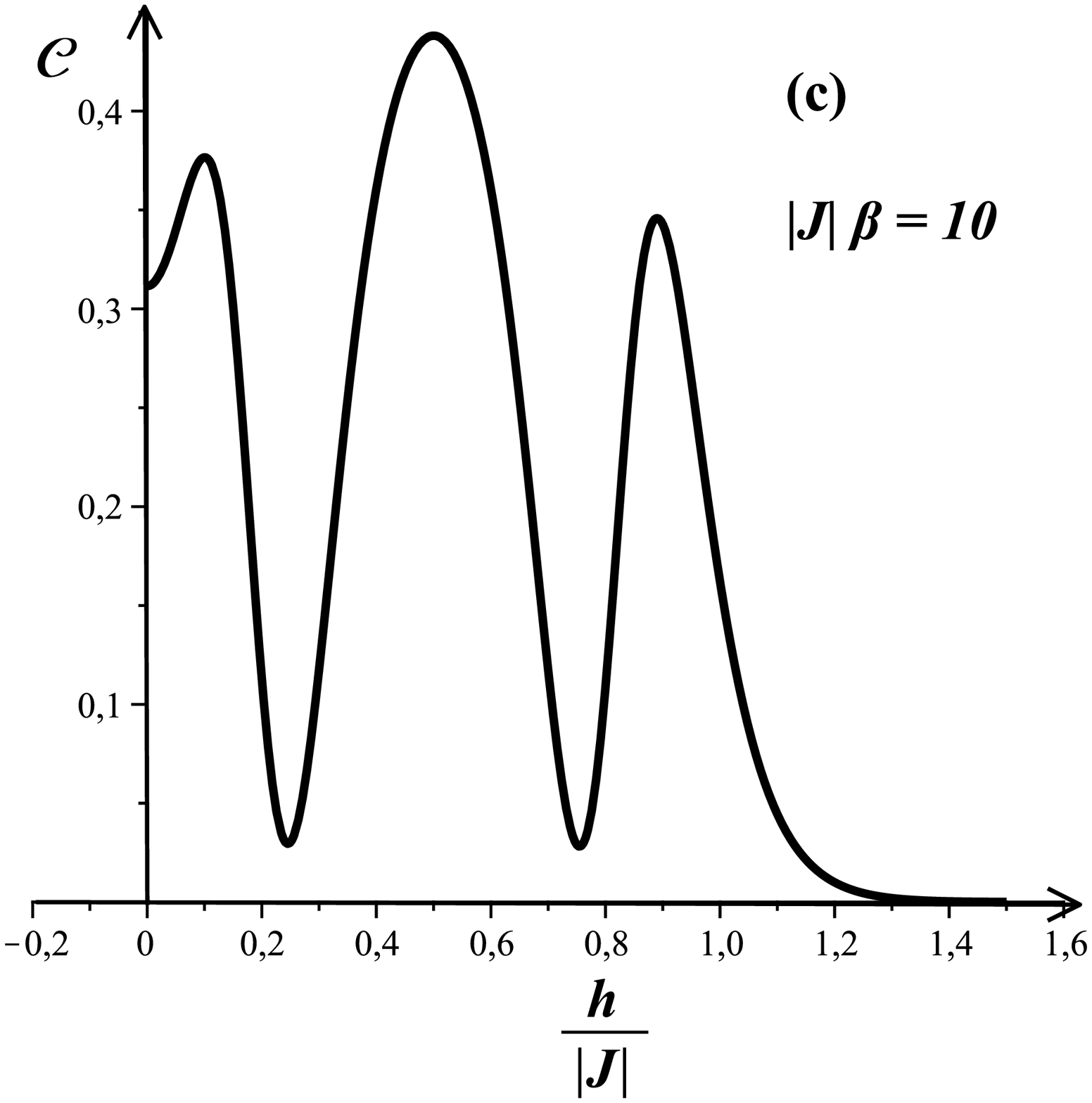}
\end{center}
\vspace{-0.7cm}
\caption{
The specific heat per site $\cal C$ of the ferromagnetic BEG 
model as a function of the external magnetic field $\frac{h}{|J|}$, 
with: $J= -1$, $\frac{K}{|J|} = -1.5$ and $\frac{D}{|J|} = 0.25$. 
In ($a$) and ($b$) we take $|J| \beta = 10^3$ and consider the 
variable $\frac{D}{|J|}$ in two intervals: $[0.23, 0.27]$
and $[0.73, 0.77]$, respectively. 
Off these intervals the specific heat per 
site is null at this temperature. In ($c$), the function
${\cal C} (\frac{h}{|J|})$ is plotted at $|J| \beta = 10$.
}   \label{fig_5}   
\end{figure}


\begin{figure} 
\begin{center}
	\includegraphics[scale=0.35, angle=180]{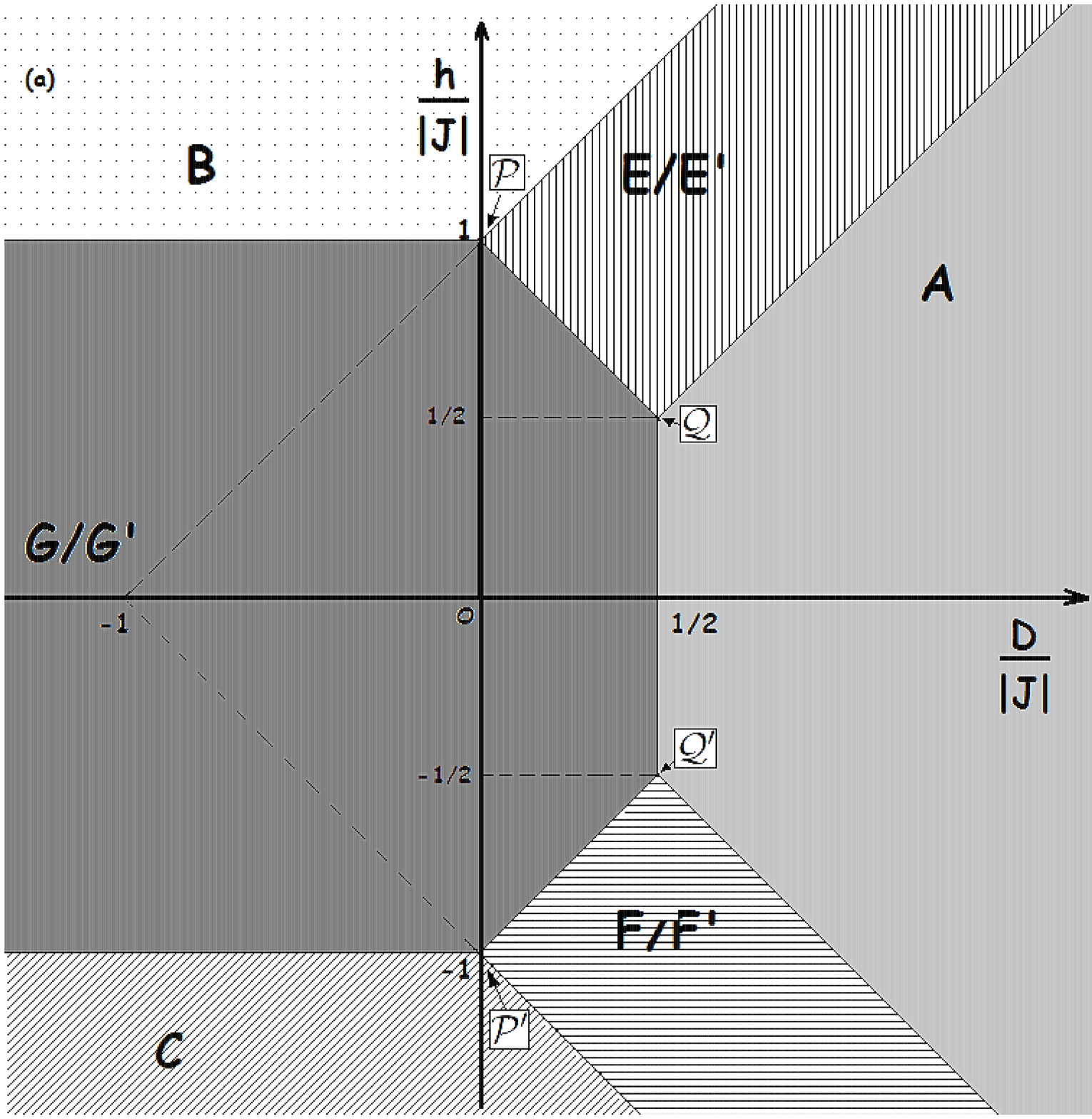}
	\ \hspace*{25mm}
	\includegraphics[scale=0.335, angle=180]{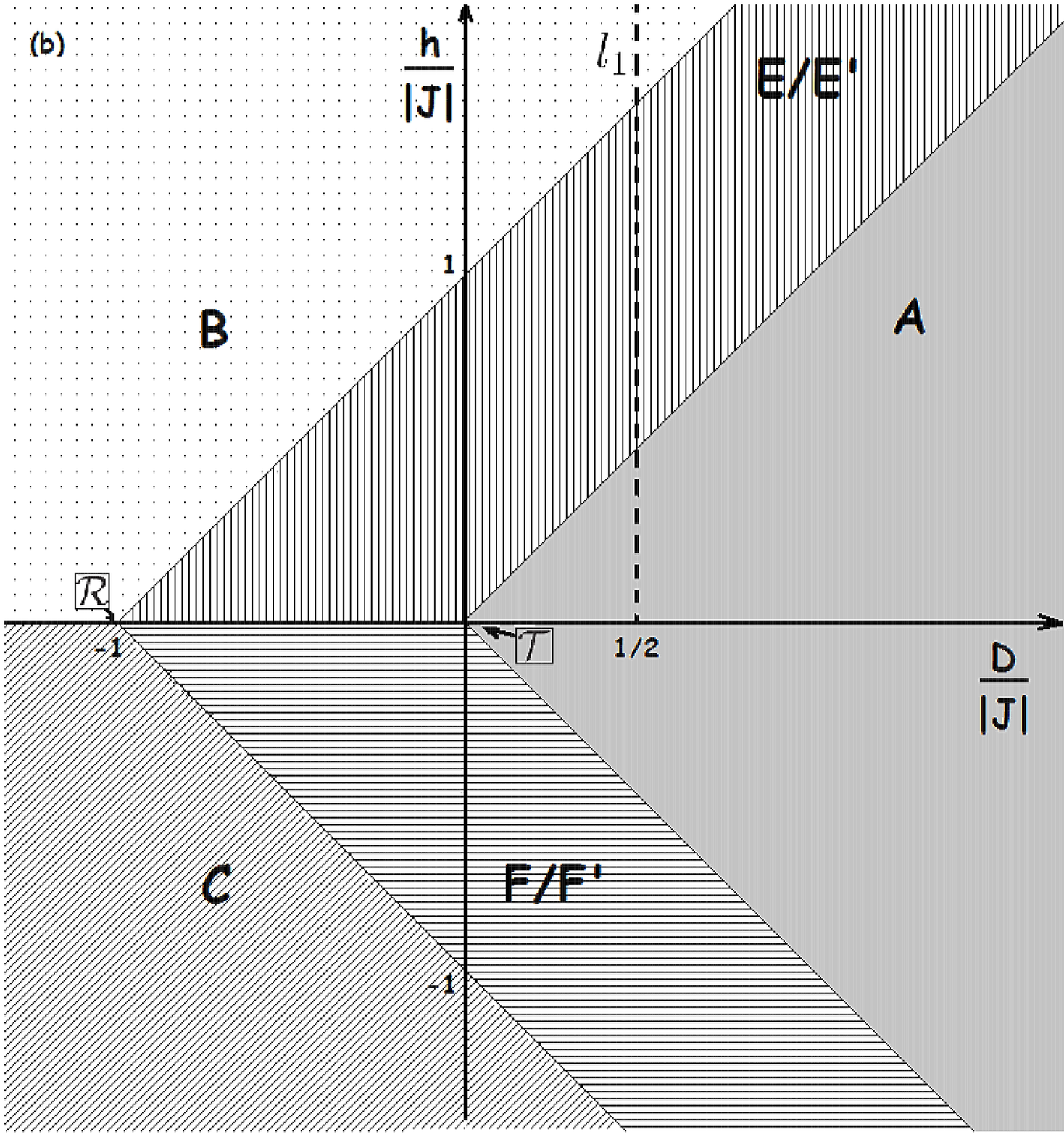}
	\end{center}
\vspace{0.7cm}
\caption{
In $(a)$, the phase diagram  of the spin-1 AF Ising model,
with single-ion anisotropy term, in the presence of a longitudinal
external magnetic field\cite{JMMM2014}, at $T=0$. 
In $(b)$, the phase diagram \ref{fig_1}a of the ferromagnetic 
BEG model, with $\frac{K}{|J|} = -2$, in the presence of a 
longitudinal magnetic field at $T=0$.
}   \label{fig_6}   
\end{figure}


\begin{figure} 
\begin{center}
\includegraphics[scale=0.4,angle= 0]{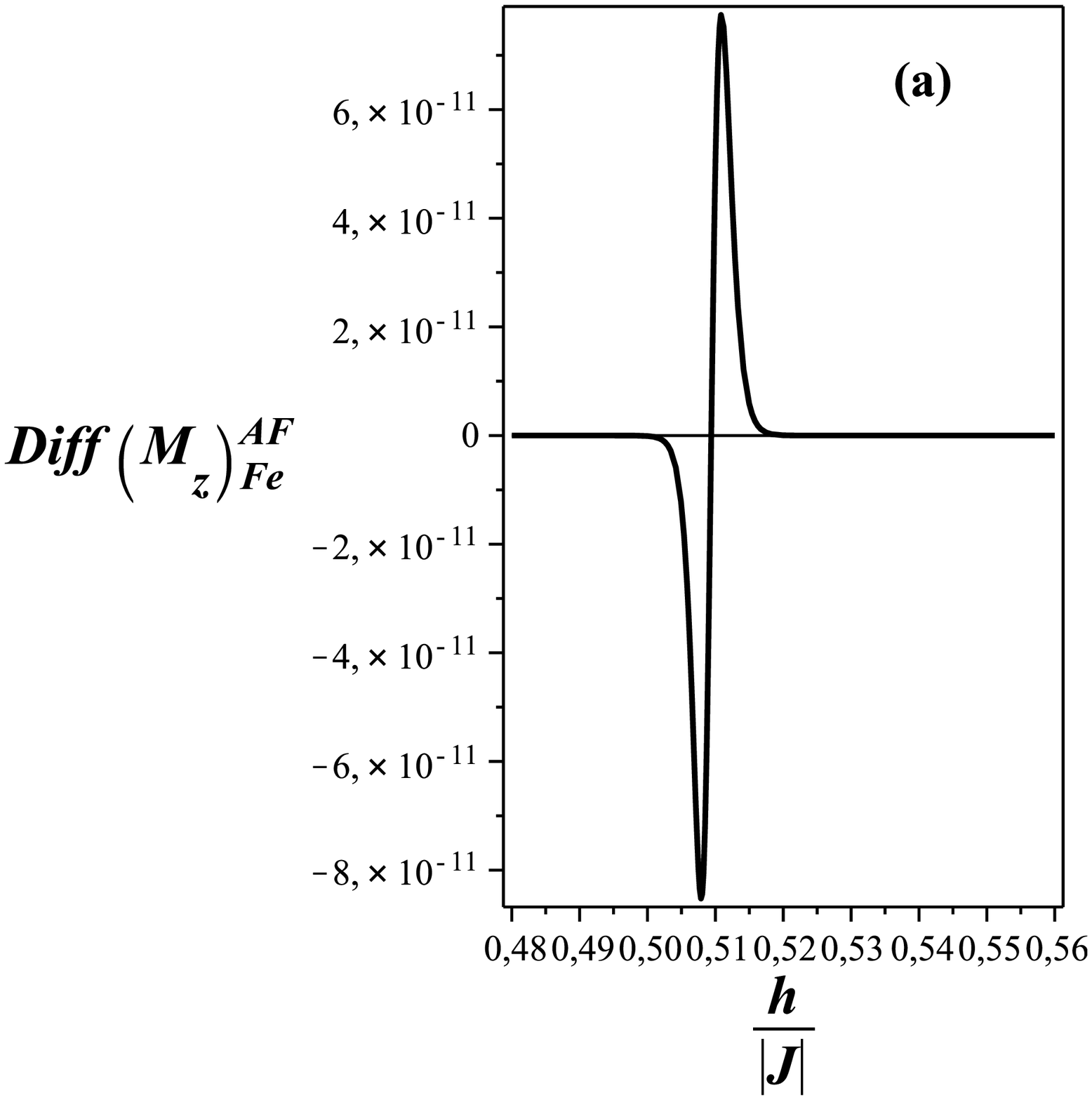}
\includegraphics[scale=0.4,angle= 0]{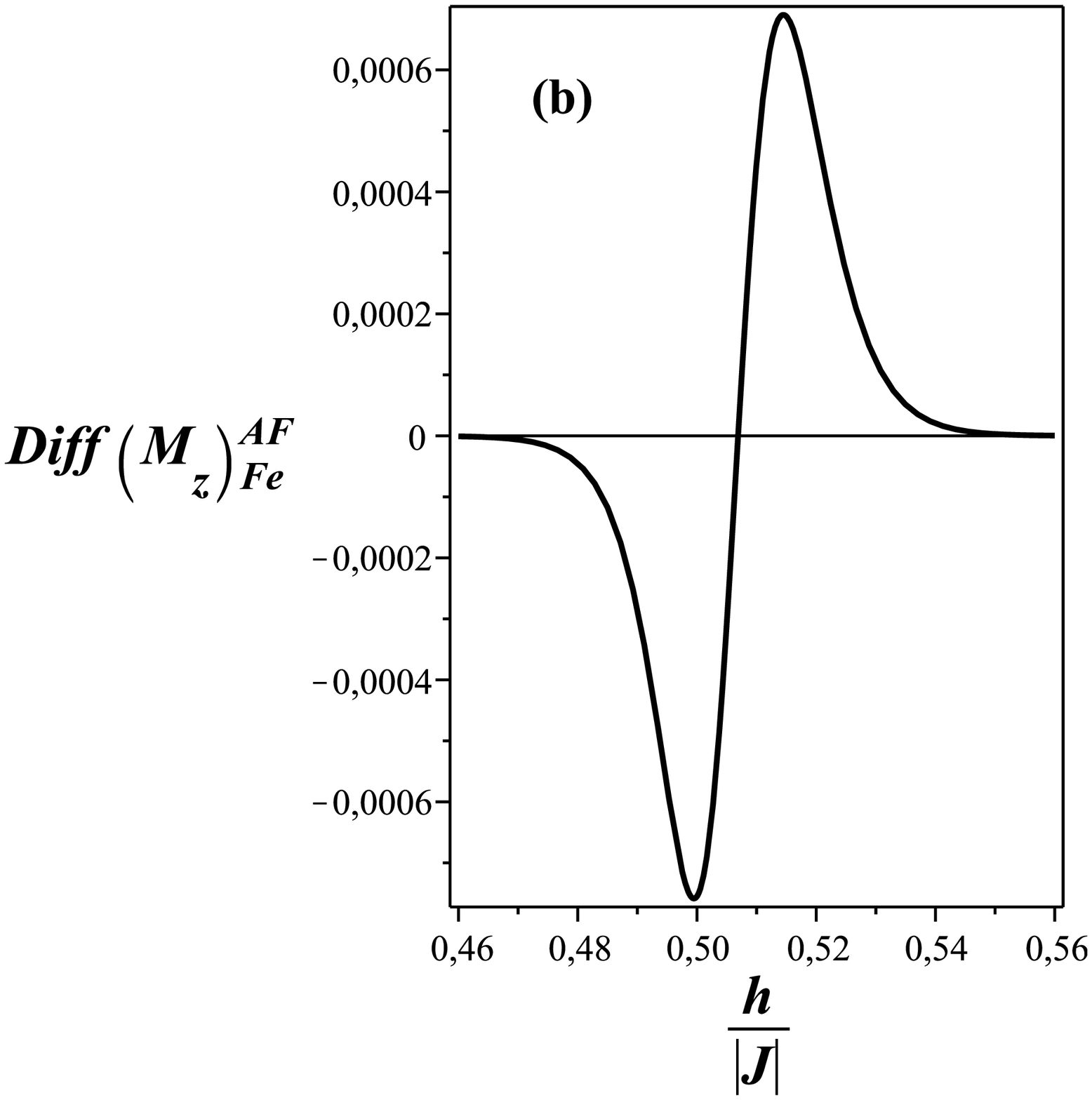}
\end{center}
\vspace{-0.7cm}
\caption{
The difference (\ref{4.7}) between the $z$-component
of the magnetization, ${\cal M}_z$, as a function 
of the external magnetic field $\frac{h}{|J|}$, of the 
ferromagnetic BEG model with $\frac{K}{|J|} = -2$,
and the spin-1 AF Ising model, with single-ion anisotropy term. 
In $(a)$ that difference is shown for $|J| \beta =500$; 
in $(b)$, for $|J| \beta = 100$.
}   \label{fig_7}   
\end{figure}
                                    

\begin{figure} 
\begin{center}
\includegraphics[scale=0.45,angle= 0]{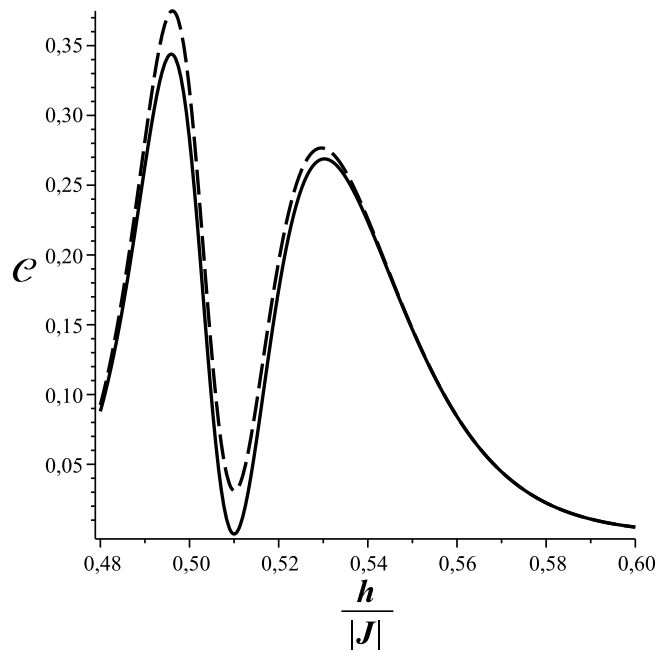}
\end{center}
\vspace{-0.7cm}
\caption{
The specific heat functions $\cal C$ per site of 
the ferromagnetic BEG model with $\frac{K}{|J|} = -2$ (solid line),
and the spin-1 AF Ising model, with single-ion 
anisotropy term (dashed line),  at $|J| \beta = 100$. Both curves 
are plotted  for $\frac{D}{|J|} = 0.51$.
}   \label{fig_8}   
\end{figure}


\begin{figure} 
\begin{center}
\includegraphics[scale=0.4,angle= 0]{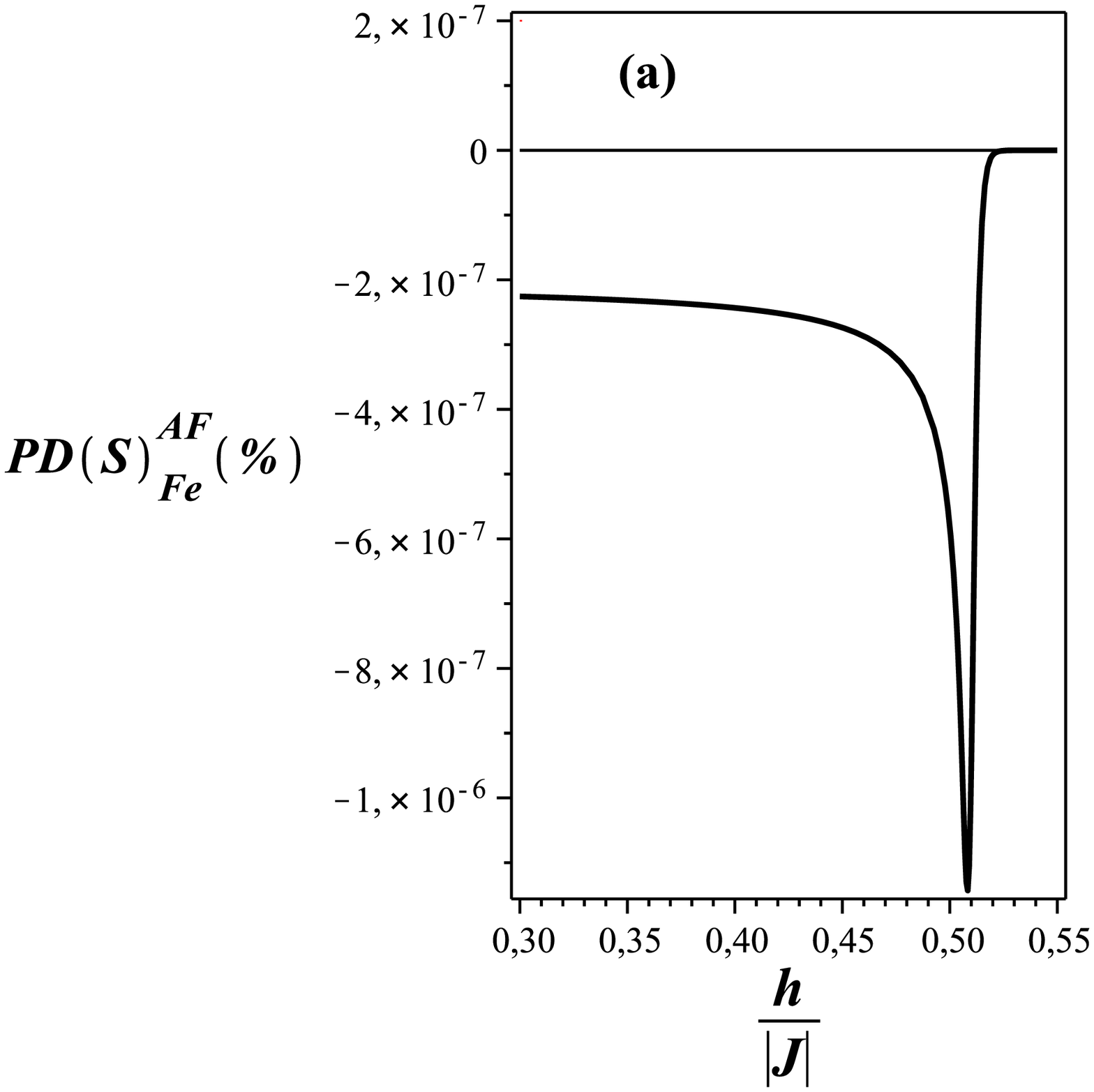}
\includegraphics[scale=0.4,angle= 0]{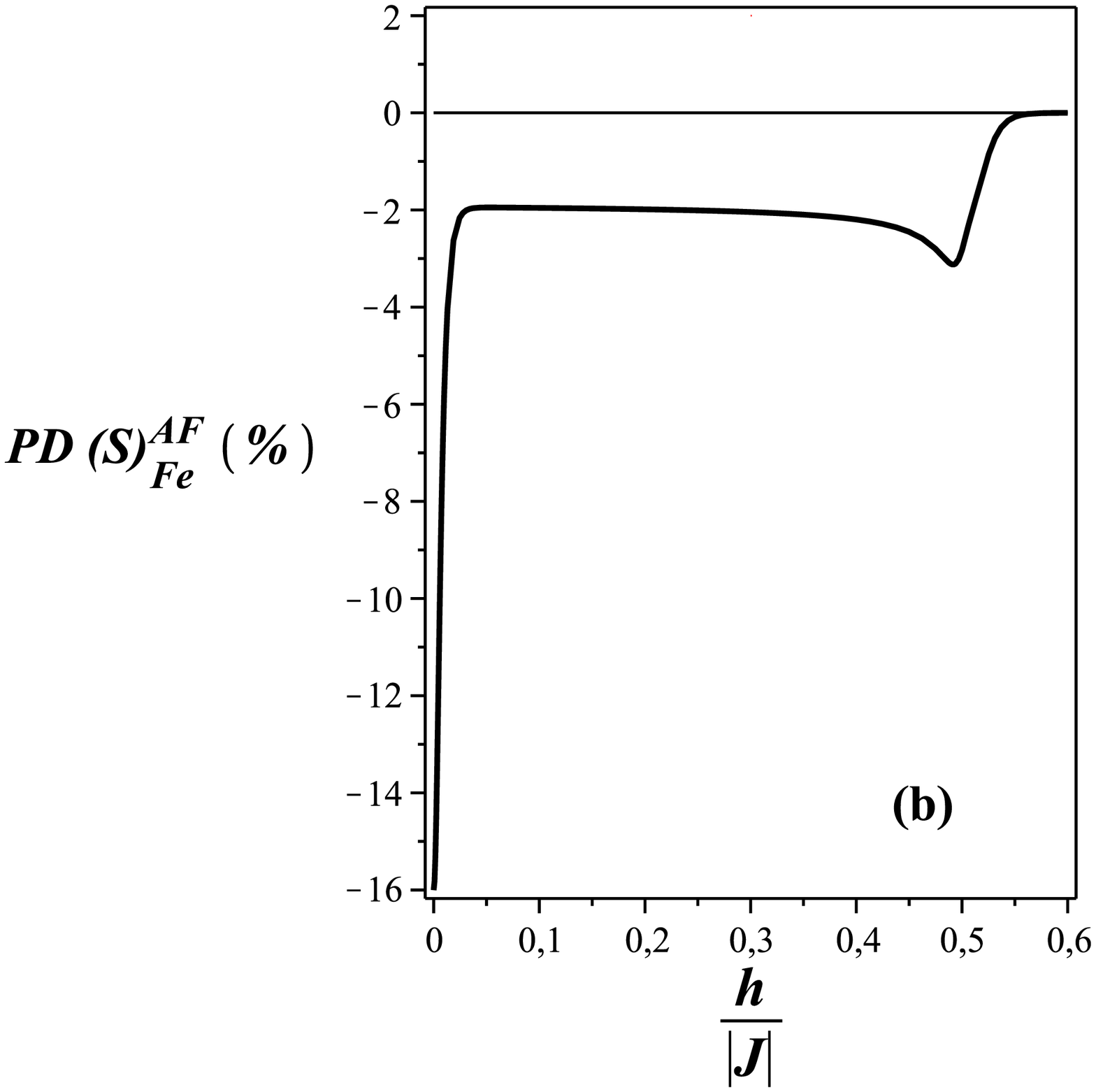}
\end{center}
\vspace{-0.7cm}
\caption{
The percent difference of the entropy per site
between the ferromagnetic BEG model with $\frac{K}{|J|} = -2$,
and the spin-1 AF Ising model, with single-ion 
anisotropy term,  with $\frac{D}{|J|}  = 0.51$.
In $(a)$, the curve is plotted at $|J| \beta = 500$; 
in $(b)$, at $|J| \beta = 100$. 
}   \label{fig_9}   
\end{figure}

\vfill


\begin{thebibliography}{}

\bibitem{simon} J. Simon {\it et al}., Nature {\bf 472}, 
04/21/2011, p. 307, and references therein.

\bibitem{kramers1} H.A. Kramers and G.H. Wannier, Phys. Rev. 
{\bf 60} (1941) 252.

\bibitem{kramers2} H.A. Kramers and G.H. Wannier, Phys. Rev. 
{\bf 60} (1941) 263.

\bibitem{baxter} R.J. Baxter. {\it Exactly Solved Models
in Statistical Mechanics.} Academic Press (1989), section 2.1.

\bibitem{JMMM2014} S.M. de Souza and M.T. Thomaz, J. of Mag. and 
Mag. Mat  {\bf 354} (2014) 205.

\bibitem{blume} M. Blume, V.J. Emery and R.B. Griffiths, Phys. 
Rev. {\bf A 4} (1971) 1071.

\bibitem{blume1966} M. Blume, Phys. Rev. {\bf 141} (1966) 517.

\bibitem{capel} H.W. Capel, Physica (Utrecht) {\bf 32} (1966) 966;
{\bf 33} (1967) 295.

\bibitem{rys} J. Bernasconi  and F. Rys, Phys. Rev. {\bf B 9} 
(1971) 3045.

\bibitem{krinsky} S. Krinsky and D. Furman, Phys. Rev. {\bf B11}
(1975) 2602. 

\bibitem{simon93} B. Simon, {\it The Statistical Mechanics of Lattice Gases}, Vol.1, 
Princeton Univ. Press, Princeton, NJ (1993).

\bibitem{schaum} M.R. Spiegel. {\it Mathematical Handbook of
Formula and Tables}. Schaum's Outline Series, Singapore (1990),
page 32.

\bibitem{reif} F. Reif, {\it Statistical Thermal Physics}. 
Mc Graw-Hill Kogakusha Ltda, International Student Edition,
Tokyo (1965), section 3.3.


\end{thebibliography}
\end{document}